\documentclass[preprint,12pt]{elsarticle}

\usepackage{amssymb}
\usepackage{mathrsfs}
\usepackage{amsfonts}
\usepackage{amsbsy}
\usepackage{booktabs}  
\usepackage{amsthm}
\usepackage{natbib}
\usepackage{geometry}
\usepackage{amsmath}
\numberwithin{equation}{section}
\newtheorem{lem}{Lemma}[section]
\newtheorem{thm}[lem]{Theorem}
\newtheorem{condition}{Condition}

\newproof{pft}{Proof}
\newproof{os}{O'Connor-Simon upper bound}
\newproof{nota}{Remark}
\newproof{conclusion}{Conclusions}
\newproof{nota1}{Remark 1}
\newproof{nota2}{Remark 2}
\newproof{nota3}{Remark 3}
\journal{Annals of Physics}

\begin{document}

\begin{frontmatter}



\title{Asymptotics of the nucleus ground-state and single-hole nature of the bound states of the nuclei\footnote{This paper was originally submitted to Annals of Physics by Franco Capuzzi on March 28, 2013 and the reviewer was \lq\lq very positively impressed by the amount and seriousness of the work presented here" and considered it \lq\lq a welcome addition to the literature of the field, with a mathematical insight that goes well beyond what is usual in nuclear theory". However, after his long disease Franco was no more able to accomplish the requested revision nor the development of the new arguments he wanted to add. His friends and colleagues have decided to submit this paper even though in the original form, because it could have some interest among the Nuclear Physics community.
}}


\author{Franco Capuzzi}
\address{Dipartimento di Fisica, Universit\`a degli Studi di Pavia and Istituto Nazionale di Fisica Nucleare, Sezione di Pavia, I-27100 Pavia, Italy}

\begin{abstract}
We consider nuclei composed of nucleons which interact via two-body potentials decreasing exponentially at infinity. Protons and neutrons are not distinguished in order to simplify notations. The basic result is the rigorous mathematical proof that the bound eigenstates of the nuclei belong to the subspace $P \mathscr{H}$ spanned by the states of a single hole in the ground state $\psi_{0}$ of the parent nucleus with an extra nucleon, as in the uncorrelated models. This follows from the exponential decay of $\psi_{0}$ when a nucleon is very far apart from the residual nucleus. We prove that the real difference from the uncorrelated models is that $P \mathscr{H}$ has an infinite dimension and contains generalized single-hole states, distinguishable from the usual ones by the fact that one cannot assign a wave function to the hole. The bound eigenstates of the nuclei are just states of this kind. Some physical consequences are discussed, in particular the unexpected fact that the dynamical part of the single-hole Hamiltonian, although nonnull, does not affect the single-hole overlaps with the bound eigenstates. The decay of $\psi_{0}$ provides the asymptotic behaviours of many single-hole quantities, in particular the nuclear density matrix. Thus a by-product of this paper is the rigorous proof of the method developed by Van Neck, Waroquier and Heyde to calculate overlaps and separation energies.  
\end{abstract}




\end{frontmatter}


\section{Introduction}
\label{intro}
Most of the results presented in this paper follow from the rigorous proof of the behaviour of the ground state $\psi_{0}$ of a system of $A+1 \geq 3$ interacting identical particles of mass $m$ when one of them is far from the residual system. Let $E_{0}$ be the energy of $\psi_{0}$ and $H$ be the residual-system Hamiltonian, living in the Hilbert space $\mathscr{H}$ and having bound eigenstates $\varphi_{n}$ of energy $\varepsilon_{n}$. Our major aim is to prove that the bound eigenstates of an $A$-particle system belong to a common subspace of $\mathscr{H}$ and infer the consequences of this restriction.

To elucidate the subject of this paper let us comment briefly the V.W.H. method, developed by Van Neck et al. \cite{VWH} neglecting the center of mass motion of the nucleus and assuming that $\psi_{0}$ has $0^{+}$ total angular momentum and parity. Disregarding details as spin, isospin and angular-spin decomposition, let $a_{\mathbf{x}}$ be the annihilation operator of a nucleon of Cartesian coordinate $\mathbf{x}$. The surprising result of \cite{VWH} is that the mere knowledge of the one-body density matrix $K(\mathbf{x},\mathbf{x}')= (a_{\mathbf{x}'}\psi_{0},a_{\mathbf{x}}\psi_{0})$ is sufficient to determine the overlaps $(\varphi_{n},a_{\mathbf{x}}\psi_{0})$ and the related separation energies $\varepsilon_{n}-E_{0}$.

The V.W.H. approach is supported by no rigorous proof and the results themselves could be subject to some criticism. Let $P$ be the projection operator onto the subspace $P\mathscr{H}$ of $\mathscr{H}$ spanned by the normalizable states $a_{f} \psi_{0}$, where $a_{f}$ destroys a nucleon of wave-function $f$. Shortly, we shall call $P$-states the elements of $P \mathscr{H}$. The information contained in the matrix $K$ concerns objects related to $P \mathscr{H}$ only. Therefore one cannot understand how the separation energies $\varepsilon_{n}-E_{0}$ may be obtained from $K$ unless the states $\varphi_{n}$ belong to $P \mathscr{H}$. As a matter of fact this property is hidden into the basic equation of the V.W.H. method itself since  eq.(24) of \cite{VWH} has the unnoticed consequence that the states $\varphi_{n}$ belong to $P \mathscr{H}$. Following a widespread belief one can object that the property $\varphi_{n}\in P \mathscr{H}$ holds only in the independent particle model!
 s. Such belief is unfounded and this is just one of the results of this paper: the real distinction between the uncorrelated systems and the correlated ones lies in the different peculiarities of the subspace $P \mathscr{H}$. In the first case $P \mathscr{H}$ has a finite dimension and all its states are of the form $a_{f} \psi_{0}$ with $f$ normalizable (single-hole states in proper sense). In the second one the dimension is infinite and $P \mathscr{H}$ contains limits of single-hole states which cannot be represented in the form $a_{f} \psi_{0}$ with $f$ normalizable
 (generalized single-hole states).

Although the V.W.H. approach is not supported by rigorous mathematical proofs, in our opinion it is affected by no inconsistency. On the contrary it is very intriguing and hides interesting conceptual aspects, chiefly the property that the states $\varphi_{n}$ are $P$-states which is strongly characterizing and rich in noticeable consequences. The idea of proving rigorously both the V.W.H. method and the property $\varphi_{n}\in P \mathscr{H}$ from the asymptotic behaviour of $\psi_{0}$ was suggested by a paper of E.H.~Lieb and B.~Simon \cite{lieb} which determine the dominant term of the decay of the ground state, with center of mass removed, for systems of spinless particles interacting via a two-body potential of compact support. They prove that, when a particle is very far apart, $\psi_{0}$ decays exponentially to a multiple $c\varphi_{0}$ of the ground state $\varphi_{0}$ of $H$ and that the pointwise decay can be expressed also by means of a strong limit in $\mathscr{H!
 }$. The unnoticed consequence is that this is a limit of $P$-states so that, if $c$ is nonnull, also $\varphi_{0}$ is a $P$-state by completeness.

In this paper we show that the property $\varphi_{n}\in P \mathscr{H}$  can be extended to excited bound states $\varphi_{n}$ and is true for a wider class of two-body potentials which can be singular at finite distances and decrease exponentially at the infinity with arbitrary rate $R$.
This class contains the potentials commonly used in Nuclear Physics (Yukawa potential included) so that our interest is focused mainly on the nuclei. The antisymmetrization problem is taken in due account but protons and neutrons are not distinguished to simplify notations. The Coulomb interaction is necessarily disregarded, as well as the spin depending interactions (as spin-spin, spin-orbit and tensor interactions) so that one can factorize the $(A+1)$-body ground state and the eigenstates of $H$ into orbital and spin parts. This restriction has no physical ground and is imposed only by reasons of mathematical rigour due to the fact that we use results established, in 
papers of pure mathematics, only for systems of spinless particles. For sake of generality the rotational invariance of the two-body potential is not required. The orbital bound states $\varphi_{n}$ of the residual nucleus are supposed nondegenerate except for the central potentials which require a more specialized approach developed in Sect. 6.

In this paper, where the conceptual aspects are prominent, a correct treatment of the center of mass motion is necessary. This is obtained using Jacobi coordinates $\mathbf{r}$ and $\boldsymbol{\xi}$ , where $\mathbf{r}$ is the coordinate of a nucleon relatively to the the center of mass of the residual system and $\boldsymbol{\xi}=(\boldsymbol{\xi}_{1},...,\boldsymbol{\xi}_{A-1})$ is the group configuration of the \emph{internal} Jacobi coordinates. Let us denote by $L^{2}(\boldsymbol{\xi})$ the Hilbert space of the square summable functions (rather, distributions) of $\boldsymbol{\xi}$ which realizes $\mathscr{H}$. Let $\psi_{0}(\mathbf{r},\boldsymbol{\xi})$ be the orbital factor of the physical $(A+1)$-body ground state, with center of mass motion removed and normalized in $\mathbf{r}$ and $\boldsymbol{\xi}$. Details on the antisymmetrization procedure in Jacobi coordinates are given in the remark below eq.(2.6). The elements of $L^{2}(\boldsymbol{\xi})$ corresponding to !
 the functions $\psi_{0}(\mathbf{r},\boldsymbol{\xi})$ for every fixed $\mathbf{r}$ are denoted by $\psi_{0,\mathbf{r}}$ in Sects. 3 and 4 for an easier comparison with the related mathematical literature. In the other sections we use the notation $a_{\mathbf{r}}\psi_{0}$ for $\sqrt{A+1}\psi_{0,\mathbf{r}}$ in order that the results may be transferred directly to the case of center of mass motion neglected or fixed by an additional external potential. Even if a Fock-type formalism (rather different from the usual one) can be defined also in Jacobi coordinates, in this paper the symbol $a_{\mathbf{r}}\psi_{0}$ is treated as a mere notation. For reader's convenience the main results of Sect. 4 are summarized in the new notations at the beginning of Sect. 5 where their consequences of physical interest are discussed. We still denote by $P \mathscr{H}$ the subspace of $\mathscr{H}$ spanned by the vectors $a_{\mathbf{r}}\psi_{0}$ or, equivalently, by the vectors $a_{f}\psi_{0}=\i!
 nt \overline{f}\left(\mathbf{r}\right)a_{\mathbf{r}}\psi_{0} d!
 \mathbf{
r}$ for any square summable function $f(\mathbf{r})$.

Two points are essential to obtain the decay of the orbital ground state $\psi_{0}$: the exponential decrease of the potential and the O'Connor-Simon upper bound for the function $| \psi_{0}(\mathbf{r},\boldsymbol{\xi})|$ (see eq. (4.1)), established in \cite{Oc}-\cite{deift}. Most probably the O'Connor-Simon upper bound holds also in presence of spin-spin, tensor and spin-orbit interactions. If so and if these interactions decrease exponentially at infinity the results are consequently extended. The basic result is the proof of the strong limits in the direction $\hat{\mathbf{r}}$ of $\mathbf{r}$ \begin{equation}\label{eq:1.1}
              s-\lim_{r\rightarrow\infty}re^{k_{n}r}\xi_{n,\mathbf{r}}
             =d_{n}\left(\hat{\mathbf{r}}\right)\varphi_{n},\quad\varepsilon_{n}<\min\left\{ \varepsilon_{C},\varepsilon_{M}\right\} 
\end{equation}
with $k_{n}=\left(2\mu_{0}\left(\varepsilon_{n}-E_{0}\right)\right)^{\frac{1}{2}}$, $\mu_{0}=\frac{A}{A+1}m$ (reduced mass) and \begin{equation}\label{eq:1.2}
\xi_{n,\mathbf{r}}=a_{\mathbf{r}}\psi_{0}-\sum_{n'=0}^{n-1}\left(\varphi_{n'},a_{\mathbf{r}}\psi_{0}\right)\varphi_{n'}
\end{equation}
where the sum is empty for $n=0$. The upper bound for $\varepsilon_{n}$ in eq. (\ref{eq:1.1}) is the lowest value between the threshold energy 
$\varepsilon_{C}$ of the continuum of $H$ and an energy $\varepsilon_{M}>\varepsilon_{0}$ which depends on $A$ and decreases for growing rates $R$ of the potential's exponential decay. The complicated expression (5.1) of $\varepsilon_{M}$ is due to the effort of optimizing the result. This is obtained using the O'Connor-Simon inequality which is considered the \emph{best possible} isotropic upper bound for $|\psi_{0}\left(\mathbf{r},\boldsymbol{\xi}\right)|$. In Nuclear Physics the restriction due to $\varepsilon_{M}$ is not severe since its value is close to $\varepsilon_{C}$ and, for many nuclei, higher. On the basis of the explicit expression (5.2) of $d_{n}\left(\hat{\mathbf{r}}\right)$, its dependence on $\hat{\mathbf{r}}$ is effective (a crucial point in the applications) and one cannot exclude that it may vanish in a part or in the whole of the solid angle.

If $d_{n}\left(\hat{\mathbf{r}}\right)$ is not identically null, eq. (\ref{eq:1.1}) alone provides nearly all the results of Sects. 5 and 6: ($a$) The asymptotic behaviour of the overlaps $(\varphi_{n},a_{\mathbf{r}}\psi_{0})$ (the overlaps with scattering states of $H$ require the separated proof of Theor. 4.3). So this important property, lacking a rigorous proof up to now, is proved definitively even if under an energy cut-off. ($b$) The asymptotic behaviour of single-hole quantities as the (subtracted) nuclear densities, the corresponding density matrices and the hole Green's function at complex energies. ($c$) The property that the bound eigenstates $\varphi_{n}$ belong to $P \mathscr{H}$, so that the energy dependent part of the single-hole Hamiltonian defined by eq. (\ref{eq:2.10}) has no effect on the overlaps $(\varphi_{n},a_{\mathbf{r}}\psi_{0})$ (a sharp-cut difference from the corresponding Feshbach's theory for the particles \cite{Feshbach}). ($d$) The rigorous !
 justification of the V.W.H. method, given in Sect. 6.

\section{Notations, definitions and elements of the single-hole theory}
\label{cap2}

We consider nuclei of $A+1\geq 3$ identical nucleons, of mass $m$, interacting via a two-body potential $V$ satisfying
\begin{equation}\label{eq:2.1}
       V\left(\mathbf{s}\right)\in L^{p}\left( \mathbf{ s } \right)\;\textrm{for a }p\in\left[2,\infty\right]         
\end{equation}
and
\begin{equation}\label{eq:2.2}
   |V\left(\mathbf{s}\right)|\leqslant V_{0}e^{-  \frac{\protect{ \huge{s} }}{R}} \quad \forall s\equiv|\mathbf{s}|\geq R_{0}>0,
\end{equation}
where $L^{p}\left(\mathbf{s}\right)$ is the Banach space of the pth power summable functions of $\mathbf{s}$. No symmetry condition on $V$ and no restriction on the rate $R$ of the exponential decrease is imposed. Whatever $p$ may be, eqs.(\ref{eq:2.1}) and (\ref{eq:2.2}) jointly imply that $V$ belongs to $L^{1}\left(\mathbf{s}\right)\cap L^{2}\left(\mathbf{s}\right)$. Eq. (\ref{eq:2.1}) allows summable singularities at finite distance (e.g. the singularity in the origin of the Yukawa potential)  
as in the mathematical papers on similar subjects.

Let us consider the partition of the $(A+1)$-body system into two clusters consisting of a single nucleon of Cartesian coordinate $\mathbf{x_{0}}$ and a residual-system of nucleons of Cartesian coordinates $\mathbf{x}_{1},...,\mathbf{x}_{A}$. To remove the center of mass motion we use the Jacobi coordinates 
\begin{equation}\label{eq:2.3}
           \mathbf{r}=
           \mathbf{x}_{0}-\frac{1}{A}\sum_{j=1}^{A}\mathbf{x}_{j}, 
                           \quad\boldsymbol{\xi}_{i} =\mathbf{x}_{i}-\frac{1}{A-i}\sum_{j=i+1}^{A}\mathbf{x}_{j},\quad1\leq i\leq A-1,
\end{equation}
with associated reduced masses
\begin{equation}\label{eq:2.4}
\mu_{0}=\frac{A}{A+1}m,\quad\mu_{i}=\frac{A-i}{A-i+1}m,
\end{equation}
and center of mass coordinate $\mathbf{X}=\frac{1}{A+1}\sum_{n=0}^{A}\mathbf{x}_{n}=0$. Let $U\left(\mathbf{s}\right)=2\mu_{0}V\left(\mathbf{s}\right)$. The 
$(A+1)$-body Hamiltonian with the center of mass motion removed is
\begin{equation}\label{eq:2.5}
H^{(A+1)}=-\frac{\nabla_{ \mathbf{r}}^{2}}{2\mu_{0}}+H+\frac{U^{(I)} \left(\mathbf{r}, \boldsymbol{\xi} \right)}{2\mu_{0}}, 
                \quad \boldsymbol{\xi} = \{ \boldsymbol{\xi}_{1}, \ldots,  \boldsymbol{\xi}_{A-1}  \},
\end{equation}
where $H$ is the residual-system Hamiltonian, acting in the space $L^{2}\left(\boldsymbol{\xi}\right)$, and $U^{(I)}\left(\mathbf{r},\boldsymbol{\xi}\right)$ is $2 \mu_{0}$ times the interaction between the two clusters: $U^{(I)}\left(\mathbf{r},\boldsymbol{\xi}\right)=\sum_{i=1}^{A}U^{\left(i\right)}\left(\mathbf{r},\boldsymbol{\xi}\right)$ with \begin{equation}\label{eq:2.6}
                      U^{\left(i\right)}\left(\mathbf{r},\boldsymbol{\xi}\right)=
                                 U\left(\mathbf{x}_{0}-\mathbf{x}_{i}\right),
                                 \quad\mathbf{x}_{0}-\mathbf{x}_{i}
                        =\mathbf{r}-\frac{A-i}{A-i+1}\boldsymbol{\xi}_{i}+\sum_{j=1}^{i-1}\frac{1}{A-j+1}\boldsymbol{\xi}_{i}.
\end{equation}
Eq. (\ref{eq:2.6}) does not include spin depending interactions so that we can choose eigenfunctions of $H^{(A+1)}$ and $H$ factorized into orbital and spin parts. Dealing with asymptotics, we are interested in eigenfunctions of H without spin structure: the spin factor will be considered only in connection with the antisymmetrization problem.

Let $\psi_{0}\left(\mathbf{r},\boldsymbol{\xi}\right)$ be the ground state of $H^{(A+1)}$, of energy $E_{0}$. For a large class of potentials, including the ones obeying eqs. (\ref{eq:2.1}) and (\ref{eq:2.2}), Theor. XIII.46 of \cite{reed} proves that $\psi_{0}$ is nondegenerate. Therefore it is automatically a symmetric function of the Cartesian variables so that $\psi_{0}$, multiplied by an antisymmetric spin factor, yields the physical ground state. Let $\left\{ \varphi_{\lambda}\left(\boldsymbol{\xi}\right)\right\} $ be a complete orthonormal set of orbital eigenfunctions of $H$ of energies $\varepsilon_{\lambda}$ consisting of bound eigenstates $\varphi_{n}$, with growing energies $\varepsilon_{n}$, and of eigenstates of the continuous spectrum $\varphi_{\varepsilon}^{\left(\alpha\right)}$ with energies $\varepsilon$ and threshold $\varepsilon_{C}$. The index $\alpha$ summarizes the various degeneracy indices. The existence of a complete orthonormal set of eigenfunction!
 s has been proved in \cite{ikebe} for rapidly decreasing potentials. The ground state $\varphi_{0}$ is necessarily nondegenerate, for the same reasons as $\psi_{0}$, and hence it is a symmetrical function of the Cartesian coordinates $\mathbf{x}_{1},...,\mathbf{x}_{A}$. In the general treatment we assume that also the excited bound states are nondegenerate, which may be incorrect even in the case of nonspherical potentials. This assumption is made only to avoid a more comlicated formalism. For the special case of central potentials this formalism is described in Sect. 6.

\begin{nota}
When an eigenstate is degenerate or is not factorized into orbital and spin parts, the antisymmetrization is a laborious task working in Jacobi coordinates. One must express the eigenstate in Cartesian coordinates, perform the antisymmetrization in the usual way and then to come back to Jacobi coordinates by inverting the eqs. (\ref{eq:2.3}). In principle one could operate directly in Jacobi coordinates using, for every permutation of Cartesian coordinates, the set of Jacobi coordinates obtained permuting the variables $\mathbf{x}_{0},\mathbf{x}_{1},...,\mathbf{x}_{A}$ in the eqs. (\ref{eq:2.3}). In practice this is a more complicated procedure since at the end every addend must be differently expressed in terms of the Jacobi coordinates of eqs. (\ref{eq:2.3}).
\end{nota}

For any fixed $\mathbf{r}$ we denote $\psi_{0,\mathbf{r}}$ the element of $L^{2}\left(\boldsymbol{\xi}\right)$ generated by $\psi_{0}(\mathbf{r},\boldsymbol{\xi})$ and $U^{(i)}_{\mathbf{r}}$ the operator in $L^{2}\left(\boldsymbol{\xi}\right)$ which multiplies by $U^{(i)}(\mathbf{r},\boldsymbol{\xi})$. We also use the notations 

\begin{equation}\label{eq:2.7}
a_{\mathbf{r}}\psi_{0}\equiv\sqrt{A+1}\psi_{0,\mathbf{r}},\quad a_{f}\psi_{0}\equiv\int \overline{f}\left(\mathbf{r}\right)a_{\mathbf{r}}\psi_{0} d\mathbf{r},\quad\forall f\in L^{2}\left(\mathbf{r}\right),
\end{equation}
useful for a comparison with the formulae used in most papers, where the center of mass motion is disregarded or fixed by an external potential. Note that in this paper we do not use the Fock-space formalism proper to the Jacobi coordinates, which is rather different from the usual one and could give rise to misunderstandings.

Let us outline now some elements of the single-hole theory, developed in \cite{boffi}, \cite{mah} and expressed here in Jacobi coordinates. The one-body density matrix 
\begin{equation}\label{eq:2.8}
              K\left(\mathbf{r},\mathbf{r}'\right)=
               \left(a_{\mathbf{r}'}\psi_{0},a_{\mathbf{r}}\psi_{0}\right)
               =\left(A+1\right)\int\overline{\psi}_{0}\left(\mathbf{r}',\boldsymbol{\xi}\right)\psi_{0}\left(\mathbf{r},\boldsymbol{\xi}\right)d\boldsymbol{\xi}
\end{equation}
is the integral kernel of a compact self-adjoint operator in $L^{2}(\mathbf{r})$. It has an orthonormal complete set of eigenfunctions $u_{\nu}(\mathbf{r})$ (natural orbitals)
 related to discrete eigenvalues $n_{\nu}$ (occupation numbers) belonging to $[0,1]$ and satisfying $\sum_{\nu}n_{\nu}=A+1$. Let $a_{\nu} \psi_{0}$ be the vectors defined by the second eq. (\ref{eq:2.7}) for $f=u_{\nu}$, which are null if and only if $n_{\nu}=0$. The scalar products $( \varphi,a_{\mathbf{r}} \psi_{0} )$ with any $\varphi \in L^{2}(\boldsymbol{\xi})$ are called \emph{single-hole overlap functions} with $\varphi$ (in short \emph{overlaps} with $\varphi$). They are always square summable functions of $\mathbf{r}$, even if $\varphi$ is an eigenfunction of the continuum of $H$. The overlaps $( \varphi_{\lambda},a_{\mathbf{r}} \psi_{0} )$ with the eigenfunctions of the discrete and the continuous spectrum are described in a framework which parallels the Feshbach's projection approach to the nuclear reactions. Details can be found in \cite{boffi} and \cite{mah}. This frame is based on the hole projection-operators $P$ and $Q$ defined by 
\begin{equation}\label{eq:2.9}
             P\varphi=\sum_{{\nu\atop n_{\nu} \neq 0}}\left(a_{\nu}\psi_{0},\varphi\right)\frac{1}{n_{\nu}}a_{\nu}\psi_{0},\quad Q\varphi=
             \left(1-P\right)\varphi,\quad\forall\varphi\in \mathscr{H},
\end{equation}
where the series is strongly convergent and $\mathscr{H}$ is the abstract Hilbert space corresponding to $L^{2}( \boldsymbol{ \xi})$. The operator $P$ projects onto its range $P \mathscr{H}$ which is the subspace spanned by the vectors $a_{\nu} \psi_{0}$. The operator $Q$ projects onto the orthogonal complement $Q \mathscr{H}$. Shortly the vectors of these subspaces will be called \emph{$P$-} and \emph{$Q$- states}. The application of the operators $P$ and $Q$ onto the states $\varphi^{(\alpha)}_{\varepsilon}$, which do not belong to $\mathscr{H}$, gains a mathematical sense through averages weighted by test functions of $\alpha$ and $\varepsilon$. By projection techniques performed on the eigenvalue equation for $H$ one defines an energy dependent Hamiltonian $H^{(h)}(\varepsilon_{\lambda})$ sum of a static and a dynamical part defined by
\begin{equation}\label{eq:2.10}
               H_{S}^{\left(h\right)}=PHP,\quad H_{D}^{\left(h\right)}\left(\varepsilon_{\lambda}\right)=
                     w-\lim_{\delta\rightarrow + 0}PHQ\left(\varepsilon_{\lambda}-QHQ-i\delta\right)^{-1}QHP,
\end{equation}
where the limit must be understood in the weak sense. This Hamiltonian determines the $P-$ components of the eigenstates $\varphi_{\lambda}$ (and hence of the corresponding overlaps) through the equations 
\begin{equation}\label{eq:2.11}
            \left(\varepsilon_{n}-H^{\left(h\right)}\left(\varepsilon_{n}\right)\right)P\varphi_{n}=0,
            \quad\left(\varepsilon-H^{\left(h\right)}\left(\varepsilon\right)\right)P\varphi_{\varepsilon}^{\left(\alpha\right)}=
            PHQ\Omega^{\left(-\right)}\varphi_{\varepsilon}^{\left(\alpha\right)},
\end{equation}
where $\Omega^{-}$ is the partial isometry defined by the strong limit 
\begin{equation}\label{eq:2.12}
\Omega^{\left(-\right)}=s-\lim_{t\rightarrow + \infty}e^{iQHQt}e^{-iHt}
\end{equation}
performed in the subspace spanned by the scattering packets $\psi_{S}$. It transforms the eigenstates $\varphi^{(\alpha)}_{\varepsilon}$ into the eigenstates of $QHQ$ with the same eigenvalues. The operator $QHP$ is the interaction which induces the transitions from $P$-states to $Q$-states. The inhomogeneus term of the second eq. (\ref{eq:2.11}), absent in the single-particle theory of Feshbach, is due to the decay of the $P$-component of the scattering packets $\psi_{S}$: \begin{equation}\label{eq:2.13}
             s-\lim_{t\rightarrow \mp \infty}Pe^{-iHt}\psi_{S}=0.
\end{equation}
The related decay width is expressed in the stationary framework as
\begin{equation}\label{eq:2.14}
           \Gamma_{\varepsilon}^{\left(\alpha\right)}=
           \frac{2}{N_{\varepsilon}^{\left(\alpha\right)}}\textrm{Im}\left(\varphi_{\varepsilon}^{\left(\alpha\right)},
            H_{D}^{\left(h\right)}\left(\varepsilon\right)\varphi_{\varepsilon}^{\left(\alpha\right)}\right)=
           \frac{2\pi}{N_{\varepsilon}^{\left(\alpha\right)}}\sum_{\beta}\Big|\left(\Omega^{\left(-\right)}\varphi_{\varepsilon}^{\left(\beta\right)},
           QHP\varphi_{\varepsilon}^{\left(\alpha\right)}\right)\Big|^{2},
\end{equation}
where $N_{\varepsilon}^{\left(\alpha\right)}$ are the spectroscopic factors of the continuum
\begin{equation}\label{eq:2.15}
        N_{\varepsilon}^{\left(\alpha\right)}=\int\Big|\left(\varphi_{\varepsilon}^{\left(\alpha\right)},a_{\mathbf{r}}\psi_{0}\right)\Big|^{2}d\mathbf{r}.
\end{equation}

For a later comparison with the correlated systems treated here, it is useful to sketch some properties of the independent particle model where each nucleon feels an external potential $V_{ext}$ only. Let $v_{\nu}$ and $v^{( \alpha )}_{e}$ be the bound and the scattering eigenfunctions of the single-particle Hamiltonian for $V_{ext}$, having respectively energies $e_{\nu}$ and $e$. The ground state $\psi_{0}$ of $H^{(A+1)}$ is obtained from the vacuum creating $A+1$ nucleons in the states $v_{\nu}$ of lowest energy. These will be denoted by $\nu \in F$ (Fermi sea). The states $v_{\nu}$ are the natural orbitals of the density matrix $K$ with occupancies 1 for $\nu \in F$ or 0 for $\nu \notin F$.

The bound eigenstates $\varphi_{n}$ of $H$ are obtained from the vacuum creating $A$ nucleons in occupied or unoccupied orbitals $v_{\nu}$. Two types of bound states can be distinguished: $i$) Single-hole states, corresponding to $\nu \in F$ for all the nucleons. Each of them is a $P$-state with overlap coinciding with the natural orbital $v_{n}$ excluded from the Fermi sea and energy $\varepsilon_{n}=E_{0}+|e_{n}|$. $ii$) States of 1 particle-2 holes, 2 particles - 3 holes and so on, corresponding to one or more orbitals $v_{\nu}$ with $\nu \notin F$. These are $Q$-states so that the related overlaps vanish. In general the $P$-states are lower in energy then the $Q$-states but embedding of one or more $Q$-states is also possible. The ground state $\varphi_{0}$ is always a single-hole state and so belongs to $P \mathscr{H}$.

The scattering eigenstates of $H$ correspond to one or more nucleons in scattering states $v^{( \alpha )}_{e}$ and are $Q$-states with related overlaps zero. Bound eigenstates of $H$ can be embedded in its continuous spectrum. Since the $Q$-states do not contribute to the spectral decomposition of the density matrix one has
\begin{equation}\label{eq:2.16}
               K\left(\mathbf{x},\mathbf{x}'\right)=
                 \sum_{n}\left(a_{\mathbf{x}'}\psi_{0},\varphi_{n}\right)\left(\varphi_{n},a_{\mathbf{x}}\psi_{0}\right)
                  =\sum_{\nu\in F}\overline{v}_{\nu}\left(\mathbf{x}'\right)v_{\nu}\left(\mathbf{x}\right),
\end{equation}
where $\sum_{n}$ is extended only to the $\varphi_{n} \in P \mathscr{H}$. The operators $P$ and $Q$ commute with $H$ so that one has 
\begin{equation}\label{eq:2.17}
PHQ=QHP=H_{D}^{\left(h\right)}\left(\lambda\right)=0
\end{equation}
and the bound states $\varphi_{n}$ belonging to $P \mathscr{H}$ are eigenvectors of $PHP$ with eigenvalue $\varepsilon_{n}$. Likewise, the transitions from $P$-states to $Q$-states are forbidden and the decay width of eq.(\ref{eq:2.14}) is zero.

\section{Properties of the overlap functions}
\label{cap3}
In this section we deduce an integral expression of the overlaps $( \varphi_{\lambda}, \psi_{0, \mathbf{r}} )$ with the eigenfunctions $\varphi_{\lambda}$ of $H$. Hence it follows an inequality for the general overlaps $( \varphi, \psi_{0, \mathbf{r}} )$, with $\varphi \in L^{2}(\boldsymbol{\xi})$, possibly deprived of the contribution of some bound states $\varphi_{n}$. Both the results are the base for the mathematical developments of the next section.

The states involved in the overlaps $( \varphi_{\lambda}, \psi_{0, \mathbf{r}} )$ obey two general boundness conditions: there exist constants $a_{\lambda}$, $b$ and $c$ such that for every value of the variables it is
\begin{equation}\label{eq:3.1}
            |\varphi_{\lambda}(\boldsymbol{\xi})| < a_{\lambda}, \qquad
            |\psi_{0}(\mathbf{r}, \boldsymbol{\xi} )| < b e^{-c \left( r +|\boldsymbol{\xi}| \right)}
\end{equation}
The first one, obvious for the bound states, is proved for the eigenfunctions $\varphi^{(\alpha)}_{\epsilon}$ of the continuum in \cite{ikebe}. The second inequality is an easy consequence of the O'Connor-Simon upper bound for $|\psi_{0}(\mathbf{r},\boldsymbol{\xi})|$, which will be introduced and commented at the beginning of Sect. 4. To simplify the following equations the degeneracy index $\alpha$ will be omitted and the related sums will be understood in the integrals over $\varepsilon$.
\begin{thm}
For every bound or scattering eigenfunction $\varphi_{\lambda}$ of $H$ one has
\begin{equation}\label{eq:3.2}
         \left(\varphi_{\lambda},\psi_{0,\mathbf{r}}\right)=
          \frac{-1}{4\pi}\int\frac{e^{-k_{\lambda}|\mathbf{r}-\mathbf{r}'|}}{|\mathbf{r}-\mathbf{r}'|}
              \left(\varphi_{\lambda},U_{\mathbf{r'}}^{\left(I\right)}\psi_{0,\mathbf{r'}}\right)d\mathbf{r'},
            \quad k_{\lambda}\equiv\left(2\mu_{0}\left(\varepsilon_{\lambda}-E_{0}\right)\right)^{\frac{1}{2}}.
\end{equation}
\end{thm}
\begin{pft}
Let $\hat{\psi}_{0,\mathbf{q}}\left(\boldsymbol{\xi}\right)$ and $\hat{f}_{0,\mathbf{q}}\left(\boldsymbol{\xi}\right)$ be the Fourier transforms, in the variable $\mathbf{r}$, of $\psi_{0,\mathbf{r}}(\boldsymbol{\xi})$ and $f_{\mathbf{r}}\left(\boldsymbol{\xi}\right)\equiv U_{\mathbf{r}}^{\left(I\right)}\left(\boldsymbol{\xi}\right)\psi_{0,\mathbf{r}}\left(\boldsymbol{\xi}\right)$, respectively. From the eigenvalue equation for $\psi_{0}$
\begin{equation}\label{eq:3.3}
\nabla^{2}_{ \mathbf{r}}\psi_{0,\mathbf{r}}=2\mu_{0}\left(H-E_{0}\right)\psi_{0,\mathbf{r}}+f_{\mathbf{r}}
\end{equation}
one has
\begin{equation}\label{eq:3.4}
\left(H-E_{0}+q^{2}\right)\hat{\psi}_{0,\mathbf{q}}=-\hat{f}_{\mathbf{q}}
\end{equation}
and then 
\begin{equation}\label{eq:3.5}
\left(\varphi_{\lambda},\hat{\psi}_{0,\mathbf{q}}\right)=-\frac{1}{q^{2}+k_{\lambda}^{2}}\left(\varphi_{\lambda},\hat{f}_{\mathbf{q}}\right).
\end{equation}
Both the scalar products involved in eq. (\ref{eq:3.5}) are expressed by successive integrals, the inner one over $\mathbf{r}$ and the outer one over 
$\boldsymbol{\xi}$. They can be exchanged, owing to the Fubini theorem, since the integrals over $\mathbf{r}$ of the absolute values are summable over 
$\boldsymbol{\xi}$. For $\left(\varphi_{\lambda},\hat{\psi}_{0,\mathbf{q}}\right)$ this follows trivially from the eqs.(\ref{eq:3.1}). In the case of $\left(\varphi_{\lambda},\hat{f}_{\mathbf{q}}\right)$, referring to the component $U^{(1)}$ of $U^{(I)}$ for brevity, the eqs.(\ref{eq:2.6}) and the second eq.(\ref{eq:3.1}) yield
\begin{equation}\label{eq:3.6}
             \int \Big|U^{\left(1\right)}\left(\mathbf{r},\boldsymbol{\xi}\right)\psi_{0}\left(\mathbf{r},\boldsymbol{\xi} \right)\Big| d\mathbf{r} 
             \leq2\mu_{0}be^{-c|\boldsymbol{\xi}|}\int \Big|V\left(\mathbf{r}-\frac{A-1}{A} \boldsymbol{\xi}_{1}\right)\Big|e^{-cr} d\mathbf{r},
\end{equation}
where the right-hand side is summable in $\boldsymbol{\xi}$ since the convolution integral is a bounded function of $\boldsymbol{\xi}_{1}$. This follows from the Young theorem applied to $V(\mathbf{s}) \in L^{1}(\mathbf{s})$, (as noticed below eq.(\ref{eq:2.2})) and $e^{-cr} \in L^{\infty}(\mathbf{r})$. Performing the exchange, $\left(\varphi_{\lambda},\hat{\psi}_{0,\mathbf{q}}\right)$ and $\left(\varphi_{\lambda},\hat{f}_{\mathbf{q}}\right)$ are proved to be the Fourier transforms of $(\varphi_{\lambda},\psi_{0,\mathbf{r}})$ and $(\varphi_{\lambda},f_{\mathbf{r}})$. Thus antitransforming eq.(\ref{eq:3.5}) one obtains eq.(\ref{eq:3.2}).
\qedhere
\qedsymbol
\end{pft}

Eq.(\ref{eq:3.2}) is the tool to obtain an useful inequality for the overlaps $(\varphi,\psi_{0, \mathbf{r}})$ with every $ \varphi \in \L^{2} (\boldsymbol{\xi})$. Let us insert the completeness relation for the system $ \{ \varphi_{\lambda} \}$ into the scalar product $(\varphi,\psi_{0, \mathbf{r}})$. Separating the first $n$ terms involving bound states (no term for $n=0$) and using eq.(\ref{eq:3.2}) in the residual ones, one has for any $\mathbf{r}$
\begin{equation}\label{eq:3.7}
           \left(\varphi,\psi_{0,\mathbf{r}}\right)-\sum_{n'=0}^{n-1}\left(\varphi,\varphi_{n'}\right)\left(\varphi_{n'},\psi_{0,\mathbf{r}}\right)=
           \sum_{{\lambda\atop n'\geq n}}\int A_{\lambda}\left(\mathbf{r} ,\mathbf{r'}\right) d\mathbf{r'},
\end{equation}
where $\sum_{{\lambda\atop n'\geq n}}$ means $\sum_{n'=n}^{\infty}+\int_{\varepsilon_{C}}^{\infty}d\varepsilon$, $\varepsilon_{C}$ is the threshold of the continuum and
\begin{equation}\label{eq:3.8}
            A_{\lambda}\left(\mathbf{r},\mathbf{r'}\right)=
              \frac{-1}{4\pi}\frac{e^{-k_{\lambda}|\mathbf{r}-\mathbf{r}'|}}{|\mathbf{r}-\mathbf{r}'|}\left(\varphi,\varphi_{\lambda}\right)
               \left(\varphi_{\lambda},U_{\mathbf{r'}}^{\left(I\right)}\psi_{0,\mathbf{r'}}\right), \quad \lambda=n' \textrm{ or } \varepsilon.
\end{equation}
Let us settle $k_{C}=\left(2\mu_{0}\left(\varepsilon_{C}-E_{0}\right)\right)^{\frac{1}{2}}$ and $\tilde{k}=\min\left\{ k_{n},k_{C}\right\} $, understanding that the minimum is simply $k_{C}$ if $\varepsilon_{n-1}$ is the highest discrete eigenvalue of $H$. In eq.(\ref{eq:3.8}) one has $e^{-k_{\lambda}|\mathbf{r}-\mathbf{r}'|}\leq e^{-\tilde{k}|\mathbf{r}-\mathbf{r}'|}$ and the following inequalities hold:
\begin{alignat}{1}\label{eq:3.9}
\sum_{{\lambda\atop n'\geq n}}\Big|A_{\lambda}\left(\mathbf{r},\mathbf{r'}\right)\Big| & \leq\frac{1}{4\pi}\frac{e^{-\tilde{k}|\mathbf{r}-\mathbf{r}'|}}{|\mathbf{r}-\mathbf{r}'|}\sum_{{\lambda\atop n'\geq n}}\Big|\left(\varphi,\varphi_{\lambda}\right)\left(\varphi_{\lambda},U_{\mathbf{r'}}^{\left(I\right)}\psi_{0,\mathbf{r'}}\right)\Big|\nonumber\\
 & \leq\frac{1}{4\pi}\frac{e^{-\tilde{k}|\mathbf{r}-\mathbf{r}'|}}{|\mathbf{r}-\mathbf{r}'|}\left(\sum_{{\lambda\atop n'\geq n}}\Big|\left(\varphi,\varphi_{\lambda}\right)\Big|^{2}\right)^{\frac{1}{2}}\left(\sum_{{\lambda\atop n'\geq n}}\Big|\left(\varphi_{\lambda},U_{\mathbf{r'}}^{\left(I\right)}\psi_{0,\mathbf{r'}}\right)\Big|^{2}\right)^{\frac{1}{2}}\nonumber\\
 & \leq\frac{1}{4\pi}\frac{e^{-\tilde{k}|\mathbf{r}-\mathbf{r}'|}}{|\mathbf{r}-\mathbf{r}'|}||\varphi|| \; ||U_{\mathbf{r'}}^{\left(I\right)}\psi_{0,\mathbf{r'}}||,\end{alignat}
where in the second step one uses the Schwarz inequality for the Hilbert space of the ordered pairs of elements of $l^{2}$ and $L^{2}$ (in the present case the sequence $ \{( \varphi_{n'},\varphi ) \} \in l^{2}$ and the function $( \varphi_{\epsilon},\varphi ) \in L^{2}$). The third step is simply the Bessel inequality. From eqs.(\ref{eq:3.7}) and (\ref{eq:3.9}) one has for every $\varphi \in L^{2}$
\begin{alignat}{1}\label{eq:3.10}
 & \Big|\left(\varphi,\psi_{0,\mathbf{r}}\right)-\sum_{n'=0}^{n-1}\left(\varphi,\varphi_{n'}\right)\left(\varphi_{n'},\psi_{0,\mathbf{r}}\right)\Big|\leq\sum_{{\lambda\atop n'\geq n}}\int \Big|A_{\lambda}\left(\mathbf{r},\mathbf{r'}\right)\Big| d\mathbf{r'} = \int \sum_{{\lambda\atop n'\geq n}}\Big|A_{\lambda}\left(\mathbf{r},\mathbf{r'}\right)\Big| d\mathbf{r'}\nonumber\\
 & \leq\frac{1}{4\pi}||\varphi||\int \frac{e^{-k|\mathbf{r}-\mathbf{r}'|}}{|\mathbf{r}-\mathbf{r}'|}||U_{\mathbf{r'}}^{\left(I\right)}\psi_{0,\mathbf{r'}}|| d\mathbf{r'},\quad\forall k\leq\min\left\{ k_{n},k_{C}\right\} ,\end{alignat}
with $\min\left\{ k_{n},k_{C}\right\}=k_{C}$ if $\varepsilon_{n-1}$ is the highest discrete eigenvalue of $H$. The exchange between $\sum_{{\lambda\atop n'\geq n}}$ and $\int d\mathbf{r'}$ is allowed by the fact that in eq.(\ref{eq:3.9}) the last term is summable over $\mathbf{r'}$ since $||U_{\mathbf{r'}}^{\left(I\right)}\psi_{0,\mathbf{r'}}||$ decreases exponentially (see eq.(4.16) deduced directly from the O'Connor-Simon inequality). Thus one can use the dominated convergence theorem for the sum over $n'$ and the Fubini theorem for the integral over $\varepsilon$.

The present proof of eq.(\ref{eq:3.10}) is not very general and has been presented here mainly since it is simple and sufficient for the potentials considered in this paper. Indeed, the requisite of a complete set $ \{ \varphi_{\lambda} \}$ is not necessary and the same result can be obtained using the spectral family of projection operators associated to $H$ because of its self-adjointness. The unique essential condition is that the continuous spectrum of $H$ is not singular.

\section{Asymptotic behaviour of the ground state}
\label{cap4}

In this section the behaviour of $\psi_{0}\left(\mathbf{r},\boldsymbol{\xi}\right)$ for $r\rightarrow\infty$ is established by means of strong limits in $L^{2}\left(\boldsymbol{\xi}\right)$ which determine the dominant term and the successive ones below a maximum energy of the residual system. The spin depending interactions are disregarded and $\psi_{0}$ represents the orbital factor of the ground state. All the results are obtained from eqs. (\ref{eq:3.2}) and (\ref{eq:3.10}) using the following upper bound for $|\psi_{0}\left(\mathbf{r},\boldsymbol{\xi}\right)|$.
\begin{os}
For every $\delta>0$ there exists a constant $c_{\delta}$ such that 
\begin{equation}\label{eq:4.1}
               e^{\left(1-\frac{\delta}{2}\right)k_{0}\sqrt[+]{\frac{I\left(\mathbf{r},\boldsymbol{\xi}\right)}{\mu_{0}}}}
              |\psi_{0}\left(\mathbf{r},\boldsymbol{\xi}\right)|\leq c_{\delta},\quad\forall\mathbf{r},\boldsymbol{\xi},
\end{equation}
where $\mu_{0}$, $\mu_{i}$ are the reduced masses related to the Jacobi coordinates $\left\{ \mathbf{r},\boldsymbol{\xi}_{i}\right\} $, $k_{0}=\left(2\mu_{0}\left(\varepsilon_{0}-E_{0}\right)\right)^{\frac{1}{2}}$ and $I\left(\mathbf{r},\boldsymbol{\xi}\right)$ is the momentum of inertia of the $A+1$ nucleons about their center of mass:
\begin{equation}\label{eq:4.2}
I\left(\mathbf{r},\boldsymbol{\xi}\right)= \mu_{0}r^{2}+\sum_{i=1}^{A-1}\mu_{i}\xi_{i}^{2}\;.
\end{equation}
\end{os}

\begin{nota}
In Sect. 2 we noticed that $\psi_{0}\left(\mathbf{r},\boldsymbol{\xi}\right)$ is a symmetrical function of the Cartesian coordinates $\mathbf{x}_{0},\mathbf{x}_{1},...,\mathbf{x}_{A}$ or, equivalently, it is invariant under replacement of $\mathbf{r},\boldsymbol{\xi}$ by the Jacobi coordinates $\mathbf{r}',
\boldsymbol{\xi'}$ obtained permuting  the Cartesian variables in the eqs. (\ref{eq:2.3}). It is easy to check that also $I\left(\mathbf{r},\boldsymbol{\xi}\right)$ satisfies the same property.
\end{nota}

The expression (\ref{eq:4.1}) of the upper bound follows directly from the results of O'Connor \cite{Oc} and Simon \cite{Simon}, established in terms of Cartesian coordinates restricted to the plane with center of mass coordinate $X=0$. Both the papers \cite{Oc} and \cite{Simon} treat systems of distinguishable spinless particles for a very wide class of two-body potentials which contains the here considered ones as well as long-range potentials (Coulomb interaction included). The extension to identical particles follows directly from the previous remark ( see also Appendix 2 of \cite{deift}). In our knowledge the validity of an upper bound like eq.(\ref{eq:4.1}) in presence of spin depending interactions has not been treated in the mathematical literature but it is in contrast with no physical reason. If so the results of this section can be extended consequently as long as the spin depending interactions vanish exponentially at the infinity.

Eq. (\ref{eq:4.1}) is considered the best isotropic bound for $|\psi_{0}\left(\mathbf{r},\boldsymbol{\xi}\right)|$ (\cite{Oc}, \cite{deift}), except for a possible improvement to $\delta=0$ with a polynomially limited function in front of the exponential in the atomic case \cite{hoff}. Indeed the Carmona-Simon upper bound for spinless particles \cite{carm}, which holds under similar wide conditions and is not isotropic, gives a fall-off faster in some directions. It is not used here for the following reasons: $i$) It involves the Agmon metric which is not invariant under permutations of the particles. $ii$) The explicit expression of the Agmon metric has been determined fixing the center of mass by means of a particle of infinite mass, which is suitable only for atomic systems. $iii$) For $A+1=3$, which is the only case where the Agmon metric has been determined with full regard to the motion of the center of mass, we have not obtained better results.

Two preliminary lemmas are the basis of the successive developments. The first one yields an upper bound for the product of $\psi_{0}$ by the component 
$U^{\left(1\right)}\left(\mathbf{x}_{0}-\mathbf{x}_{1}\right)=U\left(\mathbf{r}-\frac{A-1}{A}\boldsymbol{\xi_{1}}\right)$ of $U^{(I)}\left(\mathbf{r},
\boldsymbol{\xi}\right)$, which is the interaction (multiplied by $2 \mu_{0}$) between a nucleon and the residual system of $A \geq 2$ nucleons (see eq. (\ref{eq:2.6})). In the proof of this lemma the O'Connor-Simon inequality is used in the form
\begin{equation}\label{eq:4.3}
              |\psi_{0}\left(\mathbf{r},\boldsymbol{\xi}\right)|
              \leq c_{\delta}e^{-\left(1-\delta\right)k_{0}\sqrt[+]{\frac{I\left(\mathbf{r},\boldsymbol{\xi}\right)}{\mu_{0}}}}F\left(\boldsymbol{\xi}\right)
\end{equation}
with
\begin{equation}\label{eq:4.4}
       F\left(\boldsymbol{\xi}\right)
       \equiv e^{-\frac{\delta}{2}k_{0}\sqrt[+]{\frac{I\left(\mathbf{0},\boldsymbol{\xi}\right)}{\mu_{0}}}}
           \leq\prod_{j=1}^{A-1}f\left(\xi_{j}\right),\quad f\left(\xi_{j}\right)\equiv e^{-\frac{\delta}{2\sqrt{2}\left(A-1\right)}k_{0}\xi_{j}},
\end{equation}
where in eq. (\ref{eq:4.3}) we have made a partial use of the inequality $I( \mathbf{r},\boldsymbol{\xi} ) \geq I( \mathbf{0},\boldsymbol{\xi} )$. The inequality (\ref{eq:4.4}) is
obtained by the repeated use of the inequality $x^{2}+y^{2}\geq\frac{1}{2}\left(|x|+|y|\right)^{2}$. In eq. (\ref{eq:4.3}) we have extracted the factor $F\left(\boldsymbol{\xi}\right)$ in order to allow integration over any variable $\boldsymbol{\xi}_{j}$ in the next proofs. Note that eq. (\ref{eq:4.3}) is not weaker than eq. (\ref{eq:4.1}) since $\delta$ is arbitrarily small.

\begin{lem}

There exists a momentum $k_{M}>k_{0}$ such that for every positive $q<k_{M}$ it is

\begin{equation}\label{eq:4.5}
     \Big| U \left( 
                   \mathbf{r}-\frac{A-1}{A} \boldsymbol{\xi_{1}}
             \right) \psi_{0} \left( \mathbf{r},\boldsymbol{\xi} \right)
     \Big| \leq
            A_{q} e^{-q\, r} \left( 1+\Big| U \left( \mathbf{r}-\frac{A-1}{A} \boldsymbol{\xi_{1}} \right) \Big| \right) F \left( \boldsymbol{\xi} \right)
\end{equation}
for a suitable choice of the constant $A_{q}$. The explicit expression of $k_{M}$ is
\begin{equation}\label{eq:4.6}
     k_{M} = g \left( k_{0}R \right) k_{0}, \quad g \left( x \right) = \begin{cases}
     \sqrt{\frac{2A}{A-1}} & \textrm{for }0<x \leq \frac{A-1}{A+1} \sqrt{\frac{2A}{A-1}}\\
     \frac{1}{x} \left( 1+\sqrt{x^{2}-\frac{A-1}{A+1}} \right) & \textrm{for }x> \frac{A-1}{A+1} \sqrt{\frac{2A}{A-1}} \end{cases}.
\end{equation}
\end{lem}
\begin{pft}
Apart from the factor $F\left(\boldsymbol{\xi}\right)$ we are not interested in the dependence on $\boldsymbol{\xi}_{2},...,\boldsymbol{\xi}_{A-1}$. Thus, without weakening the results, we can replace eq. (\ref{eq:4.3}) by 
\begin{equation}\label{eq:4.7}
  | \psi_{0} \left( \mathbf{r},\boldsymbol{\xi} \right) | 
    \leq c_{\delta} e^{- \left( 1-\delta \right) k_{0} \sqrt{r^{2}+ \frac{\mu_{1}}{\mu_{0}} \xi_{1}^{2}}} F \left( \boldsymbol{\xi} \right),
    \textrm{ where }\frac{\mu_{1}}{\mu_{0}}= \frac{A^{2}-1}{A^{2}}.
\end{equation}
Let us assume $r \geq 2R_{0}$, where $R_{0}$ is the constant introduced in eq. (\ref{eq:2.2}) (this restriction will be removed at the end of the proof). To exploit the exponential decrease of the potential we distinguish the case $\xi_{1}\leq\frac{A}{A-1}\left(r-R_{0}\right)$ which implies $|\mathbf{r}-\frac{A-1}{A}\boldsymbol{\xi_{1}}|\geq r-\frac{A-1}{A}\xi_{1} \geq r-R_{0}\geq R_{0}$. Therefore eq. (\ref{eq:2.2}) yields, for every $a \geq k_{0}R$ and $\delta > 0$,
\begin{alignat}{1}\label{eq:4.8}
 & \frac{1}{2 \mu_{0}V_{0}} \Big| U \left( \mathbf{r}-\frac{A-1}{A} \boldsymbol{\xi_{1}} \right) \Big| 
   \leq e^{- \left(1- \delta \right) \frac{1}{R}| \mathbf{r}-\frac{A-1}{A} \boldsymbol{\xi_{1}}|}\nonumber\\
 & \leq e^{- \left(1- \delta \right) \frac{k_{0}}{a}| \mathbf{r}-\frac{A-1}{A} \boldsymbol{\xi_{1}}|}
   \leq e^{- \left(1- \delta \right) \frac{k_{0}}{a}(r-\frac{A-1}{A}\xi_{1})}.\end{alignat}
Of course the replacement of $R$ by $\frac{a}{k_{0}}$ may weaken the exponential decrease of the potential, but it offers the advantage of introducing a parameter which can be chosen to optimize the upper bound of $|U\psi_{0}|$ as follows.
Eqs. (\ref{eq:4.7}) and (\ref{eq:4.8}) yield
\begin{alignat}{1}\label{eq:4.9}
\Big|U\left(\mathbf{r}-\frac{A-1}{A}\boldsymbol{\xi_{1}}\right)\psi_{0}\left(\mathbf{r},\boldsymbol{\xi}\right)\Big|\leq c'_{\delta}e^{-\left(1-\delta\right)k_{0}A_{a}\left(r,\xi_{1}\right)},\nonumber\\
\forall a\geq k_{0}R,\quad\forall\boldsymbol{\xi_{1}} \, ;\quad\xi_{1}\leq\frac{A}{A-1}\left(r-R_{0}\right),\end{alignat}
with
\begin{equation}\label{eq:4.10}
  c'_{\delta}=2\mu_{0}V_{0}c{}_{\delta},\quad A_{a}\left(r,\xi_{1}\right)
  =\frac{1}{a}(r-\frac{A-1}{A}\xi_{1})+\sqrt{r^{2}+\frac{A^{2}-1}{A^{2}}\xi_{1}^{2}}\;.
\end{equation}
For every $r$ and $a>\sqrt{\frac{A-1}{A+1}}$ the function $A_{a}\left(r,\xi_{1}\right)$ of $\mathbf{\xi_{1}}$ admits a minimal value $\gamma\left(a\right)r$
with
\begin{equation}\label{eq:4.11}
\gamma\left(a\right)=\frac{1}{a}\left(1+\sqrt{a^{2}-\frac{A-1}{A+1}}\right)
\end{equation}
which yields 
\begin{alignat}{1}\label{eq:4.12}
    \Big| U \left( \mathbf{r}-\frac{A-1}{A} \boldsymbol{\xi_{1}} \right) \psi_{0} \left( \mathbf{r},\boldsymbol{\xi} \right) \Big|
    \leq c'_{\delta}e^{-\left(1-\delta\right)\gamma \left(a\right) k_{0} \, r} F \left( \boldsymbol{\xi} \right),\nonumber\\
    \forall\mathbf{r},\boldsymbol{\xi};\quad r \geq 2 R_{0} \textrm{ and }\xi_{1} \leq \frac{A}{A-1} \left( r-R_{0} \right).\end{alignat}
Eq. (\ref{eq:4.12}) holds for every $a$ satisfying $a \geq k_{0}R$ and $a > \sqrt{\frac{A-1}{A+1}}$ . Therefore the best upper limit in $r$ of $|U\left(\mathbf{r}-\frac{A-1}{A}\boldsymbol{\xi_{1}}\right)\psi_{0}\left(\mathbf{r},\boldsymbol{\xi}\right)|$ is obtained choosing the largest value of $\gamma\left(a\right)$ for $a$ satisfying the above conditions.

For $a>\sqrt{\frac{A-1}{A}}$ the function $\gamma\left(a\right)$ is continuous, larger than 1, grows to a maximum value $\sqrt{\frac{2A}{A-1}}$
at $a = \frac{A-1}{A+1}\sqrt{\frac{2A}{A-1}}$ and then decreases slowly tending to 1 at infinity. Thus we choose
\begin{equation}\label{eq:4.13}
 a = \frac{A-1}{A+1}\sqrt{\frac{2A}{A-1}}\textrm{ for }\frac{A-1}{A+1}\sqrt{\frac{2A}{A-1}}\geq k_{0}R,\quad a=k_{0}R\textrm{ otherwise,}
\end{equation}
both satisfying $a>\sqrt{\frac{A-1}{A+1}}$ for $A > 1$. This yields
\begin{alignat}{1}\label{eq:4.14}
  \Big| U \left( \mathbf{r}- \frac{A-1}{A} \boldsymbol{\xi_{1}} \right) \psi_{0} \left( \mathbf{r},\boldsymbol{\xi} \right) \Big|
  \leq c'_{\delta}e^{-\left(1-\delta\right) k_{M}r} F \left( \boldsymbol{\xi}\right),\nonumber\\
   \forall \mathbf{r},\boldsymbol{\xi};\quad r \geq 2 R_{0},\;\xi_{1}\leq\frac{A}{A-1} \left( r-R_{0} \right),\end{alignat}
with $k_{M}$ given by eq. (\ref{eq:4.6}).

It remains to study the case $\xi_{1}>\frac{A}{A-1}\left(r-R_{0}\right)>0$. Replacing, in eq. (\ref{eq:4.7}), $r$ and $\xi_{1}$ by the smaller quantities $r-R_{0}$ and $\frac{A}{A-1} (r-R_{0})$ one has 
\begin{alignat}{1} \label{eq:4.15}
 & \Big| U \left( \mathbf{r}-\frac{A-1}{A} \boldsymbol{\xi_{1}} \right) \psi_{0} \left( \mathbf{r},\boldsymbol{\xi} \right) \Big| 
   \leq c_{\delta} e^{-\left(1-\delta\right) \sqrt{\frac{2A}{A-1}} k_{0} \left(r-R_{0} \right)} \Big| U \left( \mathbf{r}-\frac{A-1}{A} \boldsymbol{\xi_{1}} \right)    \Big| F \left( \boldsymbol{\xi}\right)\nonumber\\
 & \qquad \qquad  \leq c_{\delta}e^{-\left(1-\delta\right) k_{M} \left(r-R_{0} \right)} \Big| U \left( \mathbf{r}-\frac{A-1}{A} \boldsymbol{\xi_{1}} \right) \Big| F \left( \boldsymbol{\xi} \right), \nonumber\\ 
 &  \qquad \qquad \qquad  \forall\mathbf{r},\boldsymbol{\xi}; \quad r > 2 R_{0} \quad \textrm{and} \quad \xi_{1} > \frac{A}{A-1}\left(r-R_{0}\right),\end{alignat}
where the last inequality follows from the fact that $\sqrt{\frac{2A}{A-1}}k_{0} \geq k_{M}$ since  $g\left(x\right)\leq\sqrt{\frac{2A}{A-1}}$. Therefore eqs. (\ref{eq:4.14}) and (\ref{eq:4.15}) yield eq. (\ref{eq:4.5}) for $r \geq 2R_{0}$. This condition is not restrictive since we are interested in limits for $r\rightarrow\infty$. However, it can be removed easily observing that for $r<2R_{0}$ eq. (\ref{eq:4.5}) is trivially true for every $q$. 
\qedsymbol
\end{pft}

\begin{nota1}
The expression (\ref{eq:4.6}) of $k_{M}$ yields the best upper bound for the left-hand side of eq. (\ref{eq:4.5}) which can be obtained from the O'Connor-Simon inequality, which is more or less the best isotropic bound for $| \psi_{0}|$. We do not believe that nonisotropic bounds may improve the results of the next theorems. Since $g(x)>1$ the momentum $k_{M}$ is larger than $k_{0}$, even for a very large $R$. It depends on $R$ also through the separation energy 
$\varepsilon_{0}-E_{0}$ which, however, is affected mainly by the depth of the potential $V$ and less by $R$. So, roughly speaking, the Lemma 4.1 shows that the rate of the exponential decrease of $|U\psi_{0}|$ is almost independent on $R$ for short range interactions and decreases slowly when $R$ becomes large. 
\end{nota1}
\begin{nota2}
The presence of $U$ in the right-hand side of eq. (\ref{eq:4.5}) reflects the possibly singular character of $U$ at finite distances and cannot be removed if $U$ is an unbounded function. However, the presence of $U$ does not worsen the estimates of the following lemma which concerns quantities relevant in the successive theorems.
\end{nota2}
\begin{lem}
Let $||U_{\mathbf{r}}^{(I)}\psi_{0,\mathbf{r}}||$ be the $L^{2}\left(\boldsymbol{\xi}\right)$ norm of $U_{\mathbf{r}}^{(I)}\left(\boldsymbol{\xi}\right)\psi_{0,\mathbf{r}}\left(\boldsymbol{\xi}\right)$. For any positive $q<k_{M}$ there exist constants $B_{q}$ and $B_{\lambda,q}$ so that
\begin{equation}\label{eq:4.16}
||U_{\mathbf{r}}^{(I)}\psi_{0,\mathbf{r}}||\leq B_{q}e^{-{q\, r}}
\end{equation}
and
\begin{equation}\label{eq:4.17}
\Big|\left(\varphi_{\lambda},U_{\mathbf{r}}^{(I)}\psi_{0,\mathbf{r}}\right)\Big|\leq B_{\lambda,q}e^{-{q\, r}}
\end{equation}
for every eigenstate $\varphi_{\lambda}\left(\xi\right)$ of the discrete or continuous spectrum of the residual system.
\end{lem}
\begin{pft}
First let us prove the inequalities for the component $U_{\mathbf{r}}^{\left(1\right)}\left(\boldsymbol{\xi}\right)$$=U\left(\mathbf{r}-\frac{A-1}{A}\boldsymbol{\xi_{1}}\right)$ of $U_{\mathbf{r}}^{(I)}\left(\boldsymbol{\xi}\right)$. Using eq. (\ref{eq:4.5}) and the inequality (\ref{eq:4.4}) one has, $\forall q<k_{M},$ 
\begin{equation} \label{eq:4.18}
    ||U_{\mathbf{r}}^{\left(1\right)}\psi_{0,\mathbf{r}}||
    \leq\sqrt{\frac{A}{A-1}}A_{q}||f||^{A-2}e^{-^{q\, r}}  
    \left[||f_{0}||^{2}+\left(f_{0}^{2}*|U|^{2}\right)\left( \mathbf{r} \right)+2\left(f_{0}^{2} *|U| \right)\left( \mathbf{r} \right)\right]^{\frac{1}{2}},
\end{equation}
where $f_{0}(\xi_{1})=f( \frac{A}{A-1} \xi_{1})$ and the convolutions involve a bounded function $(f^{2}_{0})$ and a summable one ($|U|^{2}$ or $|U|$, see the comment below eq. (\ref{eq:2.2})). Therefore by the Young theorem these convolutions are continuous bounded functions of $\mathbf{r}$. This proves eq. (\ref{eq:4.16}) for the component $U^{(1)}_{\mathbf{r}}$. Analogously, using eqs. (\ref{eq:4.5}) and (\ref{eq:4.4}) and exploiting the boundness of the eigenfunctions 
$\varphi_{\lambda}\left(\boldsymbol{\xi}\right)$ (see the first eq. (\ref{eq:3.1})) one has 
\begin{equation}\label{eq:4.19}
\Big|\left(\varphi_{\lambda},U_{\mathbf{r}}^{(1)}\psi_{0,\mathbf{r}}\right)\Big|
  \leq \small{\frac{A}{A-1}A_{q} a_{ \lambda} e^{-^{q\, r}} ||f||^{A-2} \left[||f_{0}||_{1}+\left(f_{0}*|U|\right)\left( \mathbf{r} \right)\right]},\; \forall q<k_{M},
\end{equation}
where $||f_{0}||_{1}$ is the norm in $L^{1} ( \boldsymbol{\xi}_{1})$ and the convolution is of the same type as above. Thus also eqs. (\ref{eq:4.17}) is proved for $U^{(1)}_{\mathbf{r}}$. The extension to the other components $U^{(k)}_{\mathbf{r}}(\boldsymbol{\xi})$ is simple. It is sufficient to use Jacobi coordinates $\mathbf{r}$, 
$\boldsymbol{\xi'}$ obtained exchanging $\mathbf{x_{1}}$ and $\mathbf{x_{k}}$ in the eqs. (\ref{eq:2.3}) and to employ the equality $\psi_{0}\left(\mathbf{r},\boldsymbol{\xi}\right)$$ = \psi_{0}\left(\mathbf{r},\boldsymbol{\xi'}\right)$ (see the remark below eq. (\ref{eq:4.2})). Thus eqs. (\ref{eq:4.16}) and (\ref{eq:4.17}) are proved
 using the triangle inequality of the norms and of the absolute values, respectively.
\qedsymbol
\end{pft}

The next theorem yields the asymptotic behaviour of the overlaps $\left(\varphi_{\lambda},\psi_{0,\mathbf{r}}\right)$, 
where $\varphi_{\lambda}$ is an eigenstate of the residual-nucleus Hamiltonian, of energy $\varepsilon_{\lambda}$, belonging 
to the discrete or continuous spectrum. In connection with the energies 
$\varepsilon_{ \lambda}$ we define the momenta 
$k_{\lambda}=\sqrt{2\mu_{0}\left(\varepsilon_{\lambda}-E_{0}\right)}$.

\begin{thm}
For any energy $\varepsilon_{\lambda}$ with $k_{\lambda}<k_{M}$ one has, in any direction $\widehat{\mathbf{r}}=\frac{\mathbf{r}}{r}$
\begin{equation}\label{eq:4.20}
\lim_{r\rightarrow\infty}re^{k_{\lambda}r}\left(\varphi_{\lambda},\psi_{0,\mathbf{r}}\right)=c_{\lambda}\left(\widehat{\mathbf{r}}\right)
\end{equation}
with
\begin{equation}\label{eq:4.21}
c_{\lambda}\left(\widehat{\mathbf{r}}\right)=
-\frac{1}{4\pi}\int e^{k_{\lambda}\widehat{\mathbf{r}}\cdot\mathbf{r'}}
\left(\varphi_{\lambda},U_{\mathbf{r'}}^{(I)}\psi_{0,\mathbf{r}'}\right)
d\mathbf{r'}.
\end{equation} 
\end{thm}
\begin{pft}
Setting
\begin{equation}\label{eq:4.22}
f_{\lambda}\left(\mathbf{r},\mathbf{r'}\right)
\equiv\frac{r}{|\mathbf{r}-\mathbf{r'}|}
e^{k_{\lambda}\left(r-|\mathbf{r}-\mathbf{r'}|\right)}
\left(\varphi_{\lambda},U_{\mathbf{r'}}^{(I)}\psi_{0,\mathbf{r}'}\right),
\end{equation}
eq. (\ref{eq:3.2}) yields
\begin{equation}\label{eq:4.23}
re^{k_{\lambda}r}\left(\varphi_{\lambda},\psi_{0,\mathbf{r}}\right)=
-\frac{1}{4\pi}\int f_{\lambda}\left(\mathbf{r},\mathbf{r'}\right)
d\mathbf{r'}.
\end{equation}
By hypothesis $k_{\lambda}<k_{M}$. Let us choose a $q \in (k_{\lambda},k_{M})$ in order that $q - k_{\lambda}>0$ and that we can use eq. (\ref{eq:4.17}) in eq. (\ref{eq:4.22}). Exploiting also the triangle inequality $r-|\mathbf{r}-\mathbf{r'}|\leq r'$ in the exponential, one has 
\begin{equation}\label{eq:4.24}
|f_{\lambda}(\mathbf{r},\mathbf{r'})|\leq
B_{\lambda,q}\frac{r}{|\mathbf{r}-\mathbf{r'}|}e^{-\left(q-k_{\lambda}\right)r'},\quad q - k_{\lambda}>0.
\end{equation}
Let us restrict the domain of integration of eq. (\ref{eq:4.23}) to the disk 
$D_{r_{0}}(\mathbf{r} )$ $=\small{ \left\{\mathbf{r'};|\mathbf{r}-\mathbf{r'}|<r_{0} \right\} } $,
with an arbitrary radius $r_{0}$. Using the inequalities (\ref{eq:4.24}) and $r'\geq r-| \mathbf{r} - \mathbf{r}' | > r-r_{0}$ one has
\begin{equation}\label{eq:4.25}
\int_{D_{r_{0} (\mathbf{r} )}} |f_{\lambda}\left(\mathbf{r},\mathbf{r'}\right)|
d \mathbf{r'} \leq 2 \pi r_{0}^{2} B_{\lambda,q}r \,e^{-\left(q-k_{\lambda}\right) \left(r-r_{0} \right)},
\end{equation}
with $q - k_{\lambda} >0$ so that the integral tends to $0$ for $r \rightarrow \infty$. 
Therefore we can perform the limit for $r \rightarrow \infty$ of eq. (\ref{eq:4.23}) restricting the integral to the domain
 complementary to $D_{r_{0}} \left( \mathbf{r} \right)$, which has as characteristic function the step function
 $\vartheta \left( |\mathbf{r} - \mathbf{r'}|-r_{0} \right)$:
\begin{equation}\label{eq:4.26}
\lim _{r \rightarrow \infty} re^{k_{\lambda}r}\left(\varphi_{\lambda},\psi_{0,\mathbf{r}}\right)=
- \frac{1}{4\pi} \lim _{r \rightarrow \infty} \int f_{\lambda}\left(\mathbf{r},\mathbf{r'}\right)
\vartheta \left( |\mathbf{r} - \mathbf{r'}|-r_{0} \right) d \mathbf{r'}.
\end{equation}
Using the inequalities $r \leq |\mathbf{r}-\mathbf{r'}|+r'$ and $|\mathbf{r}-\mathbf{r'}| \geq r_{0}$ in
 eq. (\ref{eq:4.24}) one has
\begin{equation}\label{eq:4.27}
|f_{\lambda}(\mathbf{r},\mathbf{r'})| \vartheta \left( |\mathbf{r} - \mathbf{r'}|-r_{0} \right) \leq
B_{\lambda,q} \left( 1+\frac{r'}{r_{0}} \right) e^{-\left(q-k_{\lambda}\right)r'},
\end{equation}
where $q > k_{\lambda}$ so that the second member is a summable function of $r'$, independent on $r$. Thus by dominated convergence the limit for 
 $r \rightarrow \infty$ of the integral of eq. (\ref{eq:4.26}) can be performed under the integral sign. Since 
\begin{equation}\label{eq:4.28}
 \lim _{r \rightarrow \infty} f_{\lambda}(\mathbf{r},\mathbf{r'}) \vartheta \left( |\mathbf{r} - \mathbf{r'}|-r_{0} \right) =
e^{k_{\lambda}\widehat{\mathbf{r}}\cdot\mathbf{r'}}
\left(\varphi_{\lambda},U_{\mathbf{r'}}^{(I)}\psi_{0,\mathbf{r}'}\right),
\end{equation}
one obtains eqs. (\ref{eq:4.20}) and (\ref{eq:4.21}).
\qedsymbol
\end{pft}

\begin{nota}
In eq. (\ref{eq:4.20}) we deal with a directional limit.The dependence of $c_{\lambda}$ on the direction $\widehat{\mathbf{r}}$ is an essential point, even if the potential is spherically symmetric (see the comment below eq. (6.7)). Nevertheless, as follows easily from eqs. (\ref{eq:4.23}), (\ref{eq:4.25}) and (\ref{eq:4.27}), the boundness property of the function involved in the limit  holds uniformly:
\begin{equation}\label{eq:4.29}
 re^{k_{\lambda}r} | \left( \varphi_{\lambda},\psi_{0,\mathbf{r}} \right) | \leq c, \; \forall \widehat{\mathbf{r}}.
\end{equation} 
\end{nota}

The following theorem yields the asymptotic behaviour of the vectors 
\begin{equation}\label{eq:4.30}
\chi_{n,\mathbf{r}}=\psi_{0,\mathbf{r}}-\sum_{n'=0}^{n-1}\left(\varphi_{n'},\psi_{0,\mathbf{r}}\right)\varphi_{n'},
\end{equation}
where we understand that for $n=0$ the sum is empty. Let us recall that $k_{C}=\sqrt{2\mu_{0}\left(\varepsilon_{C}-E_{0}\right)}$, where $\varepsilon_{C}$ is the threshold of the continuous spectrum of $H$. The symbol min denotes the minimum of a set.
\begin{thm}
For $k_{n}<\min\left\{ k_{C},k_{M}\right\} $
\begin{equation}\label{eq:4.31}
s-\lim_{r\rightarrow\infty}r\, e^{k_{n}r}\chi_{n,\mathbf{r}}=c_{n}\left(\widehat{\mathbf{r}}\right)\varphi_{n},
\end{equation} 
where $s-\lim$ denotes the strong limit in the direction $\widehat{\mathbf{r}}=\frac{\mathbf{r}}{r}$
and $c_{n}\left(\widehat{\mathbf{r}}\right)$ is the constant $c_{\lambda}\left(\widehat{\mathbf{r}}\right)$ of eq.(\ref{eq:4.21}) in the discrete case.
\end{thm} 
\begin{pft}
Let $q$ and $k$ be two arbitrary numbers satisfying
\begin{equation}\label{eq:4.32}
0<q<k<\min\left\{ k_{n+1},\, k_{C},\, k_{M}\right\} ,
\end{equation}
understanding that the minimum is simply $\min\left\{ k_{C},k_{M}\right\} $ if $\varepsilon_{n}$ is the highest discrete eigenvalue of $H$. Since $q$ and $k$ satisfy the conditions for eqs. (\ref{eq:4.16}) and (\ref{eq:3.10}) (used here with $n$ replaced by $n+1$), one has 
\begin{equation}\label{eq:4.33}
    |\left(\varphi,\chi_{n+1,\mathbf{r}}\right)|
    \leq\frac{B_{q}}{4\pi}||\varphi||\int \frac{e^{-k\,|\mathbf{r}-\mathbf{r'}|-q\, r'}}{|\mathbf{r}-\mathbf{r'}|} d\mathbf{r'},
     \;\forall\varphi\in L^{2}\left(\boldsymbol{\xi}\right).
\end{equation}
From this equation, using the triangle inequality $r'=|\mathbf{r}-(\mathbf{r}-\mathbf{r'})|\geq r-| \mathbf{r}- \mathbf{r}'|$, it follows
directly that for every $q<\min\left\{ k_{n+1},\, k_{C},\, k_{M}\right\} $ one can find a $k$ satisfying eq. (\ref{eq:4.32}) and a related constant 
\begin{equation}\label{eq:4.34}
B_{k,q}'\equiv\frac{B_{q}}{4\pi}\int \frac{e^{-(k-q)s}}{s} d\mathbf{s}
\end{equation}
such that 
\begin{equation}\label{eq:4.35}
|\left(\varphi,\chi_{n+1,\mathbf{r}}\right)|\leq B_{k,q}'||\varphi||e^{-q\, r},\;\forall\varphi\in L^{2}\left(\boldsymbol{\xi}\right).
\end{equation} 
Choosing $\varphi=\chi_{n+1,\mathbf{r}}$ one has 
\begin{equation}\label{eq:4.36}
||\chi_{n+1,\mathbf{r}}||\leq B_{k,q}'e^{-q\, r},\;\forall q<\min\left\{ k_{n+1},\, k_{C},\, k_{M}\right\} .
\end{equation} 
Since $k_{n}<\min\left\{ k_{C},k_{M}\right\} $ one can find a $q>k_{n}$ for which eq. (\ref{eq:4.36}) holds. Thus $r\, e^{k_{n}r}||\chi_{n+1,\mathbf{r}}||$ converges to zero and using eq. (\ref{eq:4.30}), with $n$ replaced by $n+1$, one has
\begin{equation}\label{eq:4.37}
s-\lim_{r\rightarrow\infty}r\, e^{k_{n}r}\chi_{n,\mathbf{r}}=\varphi_{n}\lim_{r\rightarrow\infty}r\, e^{k_{n}r}\left(\varphi_{n},\psi_{0,\mathbf{r}}\right),
\end{equation}
from which using eq. (\ref{eq:4.20}), which holds since $k_{n}<k_{M}$, eq. (\ref{eq:4.31}) follows.
\qedsymbol
\end{pft} 
\begin{nota1}
Theor.4.4 is subject to the condition $k_{n}<k_{C}$, absent in Theor.4.3 : differently from eq. (\ref{eq:4.20}), eq. (\ref{eq:4.31}) does not hold if $\varphi_{n}$ is a bound state embedded in the continuous spectrum.
\end{nota1}
\begin{nota2}
For $n=0\,$  eq. (\ref{eq:4.31}) has been proved in Theor.5.5 of \cite{lieb} considering two-body potentials of compact support. The present treatment, which combines the O'Connor-Simon inequality with eqs. (\ref{eq:3.2}) and (\ref{eq:3.10}), extends the result to a wider class of rapidly decreasing potentials and to excited bound states of the residual system. The practical interest of these extensions will be discussed at the beginning of the next section. 
\end{nota2}
\begin{nota3}
Eq. (\ref{eq:4.31}) is established in terms of strong limits so that it implies the corresponding weak limits and the convergence of the related norms: for $k_{n}<\min\left\{ k_{C},k_{M}\right\} $ one has 
\begin{equation}\label{eq:4.38}
        \lim_{r\rightarrow\infty}r\, e^{k_{n}r}\left(\varphi,\chi_{n,\mathbf{r}}\right)
        =c_{n}\left(\widehat{\mathbf{r}}\right)\left(\varphi,\varphi_{n}\right),\;
        \forall\varphi\in L^{2}\left(\boldsymbol{\xi}\right),
\end{equation}
and 
\begin{equation}\label{eq:4.39}
\lim_{r\rightarrow\infty}r\, e^{k_{n}r}||\chi_{n,\mathbf{r}}||=|c_{n}\left(\widehat{\mathbf{r}}\right)|.
\end{equation}
\end{nota3} 

The question whether eq. (\ref{eq:4.31}) holds also in pointwise sense lies outside the purposes of this paper, but it may be useful in other applications. It is interesting to note that for $n=0$ the pointwise eq. (\ref{eq:4.31}) can be deduced easily from eq. (\ref{eq:4.38}) using Theor.4.1 of \cite{lieb} which, for a wide class of potentials including the present ones, states that
\begin{equation}\label{eq:4.40}
             \lim_{r\rightarrow\infty}\frac{\psi_{0}\left(\mathbf{r},\boldsymbol{\xi'}\right)}{\psi_{0}\left(\mathbf{r},\boldsymbol{\xi}\right)}
             =\frac{\varphi_{0}\left(\boldsymbol{\xi'}\right)}{\varphi_{0}\left(\boldsymbol{\xi}\right)},
\end{equation}
uniformly for $\boldsymbol{\xi}$ and $\boldsymbol{\xi'}$ running through compact sets.
\begin{thm}
One has the pointwise convergence
\begin{equation}\label{eq:4.41}
    \lim_{r\rightarrow\infty}r\, e^{k_{0}r}\psi_{0}\left(\mathbf{r},\boldsymbol{\xi}\right)
    =c_{0}\left(\widehat{\mathbf{r}}\right)\varphi_{0}\left(\boldsymbol{\xi}\right),
\end{equation} 
uniformly for $\boldsymbol{\xi}$  running over compact sets.
\end{thm}
\begin{pft}
Let $\varphi\left(\boldsymbol{\xi'}\right)$ be a function of $L^{2}\left(\boldsymbol{\xi}'\right)$
with compact support and satisfying $\left(\varphi,\varphi_{0}\right)\neq 0$. Eq. (\ref{eq:4.40}), by the uniform convergence in $\boldsymbol{\xi}'$, yields  
\begin{equation}\label{eq:4.42}
           \lim_{r\rightarrow\infty}\frac{\psi_{0}\left(\mathbf{r},\boldsymbol{\xi}\right)}{\left(\varphi,\psi_{0,\mathbf{r}}\right)}
           =\frac{\varphi_{0}\left(\boldsymbol{\xi}\right)}{\left(\varphi,\varphi_{0}\right)},
\end{equation} 
uniformly on compact sets of $\boldsymbol{\xi}$.  The condition $k_{0}<\min\left\{ k_{C},k_{M}\right\} $ is satisfied since $k_{0}<k_{M}$, as observed in Remark 1 to Lemma 4.1. Thus eq.(\ref{eq:4.38}) yields
\begin{equation}\label{eq:4.43}
\lim_{r\rightarrow\infty}r\, e^{k_{0}r}\left(\varphi,\psi_{0,\mathbf{r}}\right)=c_{0}\left(\widehat{\mathbf{r}}\right)\left(\varphi,\varphi_{0}\right).
\end{equation}
which together with eq.(\ref{eq:4.42}) proves the theorem.
\qedsymbol
\end{pft} 

\section{Consequences of physical interest}
\label{cap5}

In this section we summarize and comment the main results of the previous section adopting the notation $a_{\mathbf{r}}\psi_{0}$ for 
$\sqrt{A+1}\psi_{0,\mathbf{r}}$, in order to help the comparison with the Fock-space formalism used when the center of mass motion is neglected. Accordingly, the constants $c_{\lambda}\left(\widehat{\mathbf{r}}\right)$ are replaced by 
$d_{\lambda}\left(\widehat{\mathbf{r}}\right)=\sqrt{A+1}c_{\lambda}\left(\widehat{\mathbf{r}}\right)$. The limits of the treatment are expressed more usefully in terms of the maximum energy $\varepsilon_{M}=E_{0}+\frac{k_{M}^{2}}{2\mu_{0}}$, with $k_{M}$ given by eq. (\ref{eq:4.6}): 
\begin{equation}\label{eq:5.1}
\small{
\varepsilon_{M}=E_{0}+g^{2}\left(k_{0}R\right)\left(\varepsilon_{0}-E_{0}\right),\quad g \left( x \right) =} 
\begin{cases}
\sqrt{\frac{2A}{A-1}} & \textrm{for }0<x\leq\frac{A-1}{A+1}\sqrt{\frac{2A}{A-1}} \\
\frac{1}{x}\left(1+\sqrt{x^{2}-\frac{A-1}{A+1}}\right) & \textrm{for }x>\frac{A-1}{A+1}\sqrt{\frac{2A}{A-1}}
\end{cases} .
\end{equation}

\begin{nota}
As noticed in the Remark 1 below Lemma 4.1 one has $\varepsilon_{0}<\varepsilon_{M}$ for every rate $R$ of the potential's exponential decrease.
\end{nota}

Let us recall that the results of Sect. 4 have been obtained neglecting the spin depending interactions so that even here $\psi_{0}$ and $\varphi_{n}$ denote the orbital factors of the corresponding physical eigenstates. If, as expected, the O'Connor-Simon upper bound for the ground state holds also in presence of spin depending interactions, all the following results can be extended to the unfactorized case.

\subsection{Asymptotic behaviour of the overlaps with eigenstates}
Let $\varphi_{\lambda}$ be a bound or unbound eigenstate, of energy $\varepsilon_{\lambda}$, of the residual-system Hamiltonian $H$. The statement of Theor.4.3. reads: for $\varepsilon_{\lambda}<\varepsilon_{M}$ one has, in any direction
$\widehat{\mathbf{r}}=\frac{\mathbf{r}}{r}$,
\begin{equation}\label{eq:5.2}
\lim_{r\rightarrow\infty} r\, e^{k_{\lambda}r}\left(\varphi_{\lambda},a_{\mathbf{r}}\psi_{0}\right)=d_{\lambda}\left(\widehat{\mathbf{r}}\right)\equiv-\frac{1}{4\pi}\int e^{k_{\lambda}\widehat{\mathbf{r}}\cdot\mathbf{r}'}\left(\varphi_{\lambda},U_{\mathbf{r}'}^{(I)}a_{\mathbf{r}'}\psi_{0}\right)
d\mathbf{r}',
\end{equation}
where $k_{\lambda}=\left(2\mu_{0}\left(\varepsilon_{\lambda}-E_{0}\right)\right)^{\frac{1}{2}}$.
The dependence of $d_{\lambda}\left(\widehat{\mathbf{r}}\right)$ on $\widehat{\mathbf{r}}$ is actual, even for central potentials, and it is a crucial point in order that the V.W.H. method may be operative (see the sentence below eq. (6.7)).

The weaker form of eq. (\ref{eq:5.2})
\begin{equation}\label{eq:5.3}
\lim_{r\rightarrow\infty}\left[\left(\varphi_{\lambda},a_{\mathbf{r}}\psi_{0}\right)-d_{\lambda}\left(\widehat{\mathbf{r}}\right)\frac{e^{-k_{\lambda}r}}{r}\right]=0
\end{equation}
has been used extensively in Nuclear Physics, for an unspecified $d_{\lambda}$. Eq. (\ref{eq:5.3}) was introduced first, for all the bound states, in the pioneer paper of T.~Berggren \cite{berg}. In that paper, where only potentials of finite range are considered, eq. (\ref{eq:3.2}) is written as a system of coupled equations for the overlaps, a procedure  which is too complicated for a rigorous proof. Moreover, in our opinion, one could escape hardly some type of energy cut-off due to the requirement that the integral which defines $d_{\lambda}\left(\widehat{\mathbf{r}}\right)$ in eq. (\ref{eq:5.2}) converges. In Nuclear Physics the restriction $\varepsilon_{\lambda}<\varepsilon_{M}$ does not seem severe if one considers  the values of $\varepsilon_{M}$ obtained from the experimental data for the separation energies $\varepsilon_{0}-E_{0}$ assuming a realistic value of the rate R of the exponential decay of $V (\mathbf{s})$ (e.g. $R=1,5$ fm corresponding to the pion mass). 
One has the following indications.

($i$) In general $k_{0}R$ satisfies the condition for the upper expression of $g(x)$ in eq. (\ref{eq:5.1}). For few nuclei the lower expression should be used, but the result is practically the same.

($ii$) If lower than the continuum threshold $\varepsilon_{C}$, the value of $\varepsilon_{M}$ is very close to it. However, for many nuclei $\varepsilon_{M}$ is rather higher then
 $\varepsilon_{C}$. For instance, considering the most studied nuclei $^{12}\textrm{C}$, $^{16}\textrm{O}$, $^{40}\textrm{Ca}$ and $^{208}\textrm{Pb}$, the values in MeV of $\varepsilon_{M}-E_{0}$, compared in the brackets with the corresponding values of $\varepsilon_{C}-E_{0}$, are respectively : $a$) 35 (27,2), 26 (22,3), 17,1 (14,7), 16,2 (14,8) (for the removal of a proton) and  $b$) 40,6 (27,4), 33,6 (23), 32 (21,4), 14,5 (14,1) (for the removal of a neutron).
 Thus there is also an indication that eq. (\ref{eq:5.2}) holds for bound states embedded in the continuum and for scattering states, which were two questionable points.

\subsection{Limits in the Hilbert space}
Let $\varphi_{n}$ be a bound eigenstate of $H$, of energy $\varepsilon_{n}$. The statement of Theor. 4.4 reads: for $\varepsilon_{n}<\min\left\{ \varepsilon_{C},\varepsilon_{M}\right\} $ the vectors 
\begin{equation}\label{eq:5.4}
  \xi_{n,\mathbf{r}}\equiv a_{\mathbf{r}}\psi_{0}-\sum_{n'=0}^{n-1}\left(\varphi_{n'},a_{\mathbf{r}}\psi_{0}\right)\varphi_{n'}
\end{equation}
admit the strong limit
\begin{equation}\label{eq:5.5}
  s-\lim_{r\rightarrow\infty}r\, e^{k_{n}r}\xi_{n,\mathbf{r}}=d_{n}\left(\widehat{\mathbf{r}}\right)\varphi_{n}.
\end{equation}
In comparison with eq. (\ref{eq:5.2}) note the further condition $\varepsilon_{n}<\varepsilon_{C}$, which excludes the bound states embedded in the continuum. Eq. (\ref{eq:5.5}) is equivalent to the pair of limits 
\begin{equation}\label{eq:5.6}
  \lim_{r\rightarrow\infty}r\, e^{k_{n}r}\left(\varphi,\xi_{n,\mathbf{r}}\right)=d_{n}\left(\widehat{\mathbf{r}}\right)\left(\varphi,\varphi_{n}\right),
\quad\forall\varphi\in \mathscr{H},
\end{equation}
\begin{equation}\label{eq:5.7}
  \lim_{r\rightarrow\infty}r^{2}e^{2k_{n}r}\rho_{n}\left(\mathbf{r}\right)=|d_{n}\left(\widehat{\mathbf{r}}\right)|^{2},
\end{equation}
where $\rho_{0}\left(\mathbf{r}\right)$ is the nuclear density, normalized to $A+1$, and the successive $\rho_{n}\left(\mathbf{r}\right)$ are the subtracted nuclear densities :
\begin{equation}\label{eq:5.8}
  \rho_{0}\left(\mathbf{r}\right)=||a_{\mathbf{r}}\psi_{0}||^{2},
  \quad\rho_{n}\left(\mathbf{r}\right)=||\xi_{n,\mathbf{r}}||^{2} =
\rho_{0}\left(\mathbf{r}\right) - \sum_{n'=0}^{n-1}|\left(\varphi_{n'},a_{\mathbf{r}}\psi_{0}\right)|^{2}.
\end{equation}

\begin{nota}
 Since $\varepsilon_{M}>\varepsilon_{0}$, even for very large rates $R$ of the potential's decrease, eqs. (\ref{eq:5.2}) and (\ref{eq:5.5})-(\ref{eq:5.7}) are always true for $n=0$. 
\end{nota}

\subsection{Asymptotic behaviour of single-hole kernels}
Let $A\left(\mathbf{r},\mathbf{r}'\right)=\left(a_{\mathbf{r}'}\psi_{0},Aa_{\mathbf{r}}\psi_{0}\right)$ be a single-hole kernel for an operator $A$ defined on 
$\varphi_{0}$ and having adjoint $A^{\dagger}$ defined on the vectors $a_{\mathbf{r}}\psi_{0}$, as well. Choosing $n=0$ and $\varphi= A^{\dagger} a_{\mathbf{r}'}\psi_{0}$ in eq. (\ref{eq:5.6}) one has directly, however large $R$ may be,
\begin{equation}\label{eq:5.9}
  \lim_{r\rightarrow\infty}r\, e^{k_{0}r}A\left(\mathbf{r},\mathbf{r}'\right)
   =d_{0}\left(\widehat{\mathbf{r}}\right)\left(a_{\mathbf{r}'}\psi_{0},A\varphi_{0} \right) .
\end{equation}
Typical examples are the density matrix $K$ (if $A$ is the identity operator), the energy weigthed density matrix ($A=H$), the one-hole Green's function at complex energies $z$ ($A=\left(z-H\right)^{-1}$) and the one-hole Hamiltonian $H^{\left(h\right)}\left(z\right)$ ($A=PHP+PHQ\left(z-H\right)^{-1}QHP$).

Under the further condition that $A$ is a bounded operator, so that $r\, e^{k_{0}r}Aa_{\mathbf{r}}\psi_{0}$ converges to 
$d_{0}\left(\widehat{\mathbf{r}}\right)A\varphi_{0}$, one has the joint limit
\begin{equation}\label{eq:5.10}
  \lim_{r\rightarrow\infty}r^{2}e^{2k_{0}r}\left(a_{\mathbf{r}}\psi_{0},Aa_{\mathbf{r}}\psi_{0}\right)
  =|d_{0}\left(\widehat{\mathbf{r}}\right)|^{2}\left(\varphi_{0},A\varphi_{0}\right).
\end{equation}
An interesting application of eq. (\ref{eq:5.10}) is the case of the one-hole Green's function at complex energies. If $A$ is an unbounded operator eq. 
(\ref{eq:5.10}) holds if $A$ is closed, provided that $s-\lim_{r\rightarrow\infty}r\, e^{k_{0}r}Aa_{\mathbf{r}}\psi_{0}$ exists.

\subsection{Hilbert subspace of the bound eigenstates}
Let us suppose that all the coefficients $d_{n}\left(\widehat{\mathbf{r}}\right)$ are different from zero, even in a single possibly different direction $\widehat{\mathbf{r}}$. In such a case eq. (\ref{eq:5.5}) implies that the bound states $\varphi_{n}$  with 
$\varepsilon_{n}<\min\left\{ \varepsilon_{C},\varepsilon_{M}\right\} $ belong to the subspace $P \mathscr{H}$ spanned by the vectors $a_{\mathbf{r}}\psi_{0}$ or, equivalently, satisfies $P \varphi_{n}=\varphi_{n}$ where $P$ is the projection operator defined in eq. (2.9). In fact, since $\xi_{0,\mathbf{r}}=a_{\mathbf{r}} \psi_{0}$, the vectors $r\, e^{k_{0}r}\xi_{0,\mathbf{r}}$ belong to $P \mathscr{H}$ so that also their strong limit $d_{0}\left(\widehat{\mathbf{r}}\right) \varphi_{0}$ belongs to $P \mathscr{H}$ due to the completeness of this subspace. If $d_{0}\left(\widehat{\mathbf{r}}\right) \neq 0$ for a single direction $\widehat{\mathbf{r}}$, the property is transferred to $\varphi_{0}$, hence to the vectors $r\, e^{k_{1}r}\xi_{1,\mathbf{r}}$, to their limit and so on. Of course, if for a given $n$ the coefficient $d_{n}\left(\widehat{\mathbf{r}}\right)$ is identically null, one cannot infer that $\varphi_{n}$ is a $P$-state and the chain of the successive implications could break up.

The property $\varphi_{n} \in P \mathscr{H}$ has an important consequence which could not be foreseen in the single-hole theory developed in \cite{boffi} and \cite{mah}. It is well-known that in the independent particle model the dynamical part $H^{(h)}_{D}$ of the single-hole Hamiltonian $H^{(h)}$, defined in eq. (\ref{eq:2.10}), is null since the projection operator $P$ commutes with $H$. This is no longer true if the nucleons interact via a two-body potential. In spite of this, due to 
$P \varphi_{n} = \varphi_{n}$, the operator $H^{(h)}_{D}(\varepsilon_{n})$ annihilates 
$\varphi_{n}$ since one has $HP \varphi_{n}=\varepsilon_{n}P \varphi_{n}$ which implies $QHP \varphi_{n}=0$. Thus the term involving $H^{(h)}_{D}$ has no relevance in determining the single-hole overlaps. This points out a remarkable difference from the single-particle theory of Feshbach \cite{Feshbach} where the dynamical part of the single-particle Hamiltonian plays an essential role in the description of the single-particle overlaps. 
As well as this, it is worthwhile remarking an analogous difference between $H^{(h)}$ and the Hamiltonian for the self-energy, also given by eq. (\ref{eq:2.10}) but with $P$ replaced by an extended projection operator (see eqs. (2.24) and (3.6) of \cite{capma}). This Hamiltonian too provides the single-hole overlaps with the eigenstates $\varphi_{n}$ through an eigenvalue equation identical to the first eq. (\ref{eq:2.11}). But in this case one has not $P \varphi_{n} = \varphi_{n}$ so that the role of the dynamical part of the self-energy becomes essential.
Of course all these conclusions are subject to the conditions 
$\varepsilon_{n}<\min\left\{ \varepsilon_{C},\varepsilon_{M}\right\} $ and $d_{n}\left(\widehat{\mathbf{r}}\right) \neq 0$ at least for a direction $\widehat{\mathbf{r}}$. In \cite{boffi} the operator $QHP$ is interpreted as the interaction responsible for the transitions from $P$-states to $Q$-states. These are all forbidden in the uncorrelated systems since the operator $QHP$ is zero. In the real systems they are allowed due to the correlation effects. The result $\varphi_{n}=P \varphi_{n}$ exhibits the unexpected exception of the transitions from the bound eigenstates $\varphi_{n}$ towards all the $Q$-states. For the unbound states 
$\varphi^{( \alpha )}_{\varepsilon}$ the situation is quite different since $QHP \varphi^{( \alpha )}_{\varepsilon}$ must be different from zero in order that the decay widths defined in eq. (\ref{eq:2.14}) do not vanish. Thus one has neither 
$ \varphi^{( \alpha )}_{\varepsilon} =P \varphi^{( \alpha )}_{\varepsilon}$ nor 
$ \varphi^{( \alpha )}_{\varepsilon} =Q \varphi^{( \alpha )}_{\varepsilon}$ and the packets of states 
$\varphi^{( \alpha )}_{\varepsilon}$ have nonnull components both in $P \mathscr{H}$ and $Q \mathscr{H}$.

\begin{nota}
The previous property of the unbound eigenstates of $H$ can affect the Hilbert nature of the bound eigenstates embedded in the continuum. Let us suppose $\varepsilon_{C} < \varepsilon_{n} < \varepsilon_{M}$ and $d_{n}\left(\widehat{\mathbf{r}}\right) \neq 0$ in a direction $\widehat{\mathbf{r}}$. The next argument follows the steps of the proof of Theor.4.4 with the only difference that here it is necessary to extract from $\xi_{n+1,\mathbf{r}}$ the contribution of the continuum up to the energy $\varepsilon_{n}$ of the bound state. In such a way one obtains  
\begin{equation}\label{eq:5.11}
s-\lim_{r\rightarrow\infty}r\, e^{k_{n}r}\left[\xi_{n,\mathbf{r}}-\sum_{\alpha}\int_{\varepsilon_{C}}^{\varepsilon_{n}}\left(\varphi_{\varepsilon}^{(\alpha)},a_{\mathbf{r}}\psi_{0}\right)\varphi_{\varepsilon}^{(\alpha)}d\varepsilon\right]=d_{n}\left(\widehat{\mathbf{r}}\right)\varphi_{n},
\end{equation}
where the integral has nonnull components both in $P \mathscr{H}$ and $Q \mathscr{H}$. Thus this property can be shared by $\varphi_{n}$. 
\end{nota}

\subsection{Comparison with the independent particle model}
In consideration of the role played by this model in the next section it is useful to compare the results of the previous subsections with the ones of the independent particle model. Of course, since the center of mass motion cannot be removed we use Cartesian coordinates $\mathbf{x}_{0}$ and $\mathbf{x}= \left\{ \mathbf{x}_{1},..., \mathbf{x}_{A} \right\} $. The operator $P$ projects onto the subspace spanned by the vectors $a_{\mathbf{x}_{0}} \psi_{0}$. 

($i$) \emph{Correlated systems}. In the normal case, i.e. when no coefficient $d_{n}\left(\widehat{\mathbf{r}}\right)$ is identically null, every bound state $\varphi_{n}$ with $\varepsilon_{n}<\min\left\{ \varepsilon_{C},\varepsilon_{M}\right\} $ is a $P$-state. The unbound states and their packets have nonnull components both in $P \mathscr{H}$ and $Q \mathscr{H}$. The same, most probably, for the bound states embedded in the continuum.

($ii$) \emph{Uncorrelated models}. As seen at the end of Sect. 2, the $A+1$ single-hole eigenstates of $H$ belong to $P \mathscr{H}$. The remaining bound or unbound eigenstates which make the system complete belong to $Q \mathscr{H}$. Thus $P \mathscr{H}$ is the subspace spanned by the single-hole eigenstates and its dimension is $A+1$.

Due to the point ($ii$), in the uncorrelated models the expansion of $a_{\mathbf{x}_{0}} \psi_{0}$ on the basis of the eigenvectors of 
$H$ is extended only to the single-hole states:
\begin{equation}\label{eq:5.12}
a_{\mathbf{x}_{0}}\psi_{0}=\sum_{n=0}^{A}\left(\varphi_{n},a_{\mathbf{x}_{0}}\psi_{0}\right)\varphi_{n},
\end{equation}
where $\varphi_{n}=a_{v_{n}} \psi_{0}$ and $v_{n}$ is the natural orbital excluded from the Fermi sea. It is an eigenfunction, belonging to the eigenvalue $e_{n}$, of the one-body Hamiltonian for the external potential $V_{\textrm{ext}}$ (which here is supposed local and obeying eqs. (\ref{eq:2.1}) and (\ref{eq:2.2})). Therefore one has
\begin{equation}\label{eq:5.13}
\left(\varphi_{n},a_{\mathbf{x}_{0}}\psi_{0}\right)=v_{n}(\mathbf{x}_{0}),\qquad\varepsilon_{n}-E_{0}=-e_{n}.
\end{equation}
By eqs. (\ref{eq:5.12}) and (\ref{eq:5.13}), the eqs. (\ref{eq:5.4}) and (\ref{eq:5.8}) become
\begin{equation}\label{eq:5.14}
\xi_{n,\mathbf{x}_{0}}=\sum_{n'=n}^{A}v_{n'}(\mathbf{x}_{0})\varphi_{n'}
\end{equation}
and
\begin{equation}\label{eq:5.15}
\rho_{n}(\mathbf{x}_{0})=||\xi_{n,\mathbf{x}_{0}}||^{2}=\sum_{n'=n}^{A}|v_{n'}(\mathbf{x}_{0})|^{2}.
\end{equation}
The asymptotic behaviour of $\xi_{n, \mathbf{x}_{0}}$ and $\rho_{n}(\mathbf{x}_{0})$ follows from the following properties of the eigenfunctions $v_{n}$. It is well-known that they satisfy the integral equation
\begin{equation}\label{eq:5.16}
v_{n}(\mathbf{x}_{0})=-\frac{1}{4\pi}\int\frac{e^{-k_{n}|\mathbf{x}_{0}-\mathbf{x}'_{0}|}}{|\mathbf{x}_{0}-\mathbf{x}'_{0}|}V_{\textrm{ext}}\left(\mathbf{x}'_{0}\right)v_{n}(\mathbf{x}'_{0})d\mathbf{x}'_{0},
\end{equation}
where $k^{2}_{n}=-2m \, e_{n}=2m(\varepsilon_{n}-E_{0})$. Moreover for every $k < k_{n}$ the functions 
$e^{k \, x_{0}}v_{n}(\mathbf{x}_{0})$ are bounded as proved below. This implies by eq. (\ref{eq:2.2}) that, however large $R$ may be, one can find a 
$q > k_{n}$ and a constant $B_{n,q}$ such that
\begin{equation}\label{eq:5.17}
|V_{\textrm{ext}}\left(\mathbf{x}{}_{0}\right)v_{n}(\mathbf{x}{}_{0})|\leq B_{n,q}e^{-qx_{0}}.
\end{equation}
The boundness of the functions $e^{k \, x_{0}}v_{n}(\mathbf{x}_{0})$ can be proved easily in the present case where $V_{\textrm{ext}}$ belongs to $L^{2}$, as observed below eq. (\ref{eq:2.2}). In fact it is sufficient to use the simple proof of Theor.VI.6 of  
\cite{Simon71} which states that for every $k < k_{n}$ the functions $e^{k \, x_{0}}v_{n}(\mathbf{x}_{0})$ belong to $L^{2}$. The same proof can be adapted to prove that this property is shared by $e^{k \, x_{0}} V_{\textrm{ext}}(\mathbf{x}_{0}) v_{n}(\mathbf{x}_{0})$. Therefore using eq. (\ref{eq:5.16}) and the inequality $x_{0} - x_{0}' \leq |\mathbf{x}_{0}-\mathbf{x}_{0}'|$one has
\begin{equation}\label{eq:5.18}
|e^{k \, x_{0}}v_{n}(\mathbf{x}_{0})| \leq 
     \frac{1}{4\pi} \int 
     \frac{e^{-(k_{n} - k)|\mathbf{x}_{0}-\mathbf{x}'_{0}|}}{|\mathbf{x}_{0}-\mathbf{x}'_{0}|}
      e^{k \, x_{0}'} \, |V_{\textrm{ext}}(\mathbf{x}_{0}') v_{n}(\mathbf{x}_{0}')| d\mathbf{x}'_{0},
\end{equation}
where the right-hand side is a convolution between two functions belonging to $L^{2}$, which is a bounded function of $\mathbf{x}_{0}$ by the Young theorem. Using eqs. (\ref{eq:5.16}) and (\ref{eq:5.17}) instead of eqs. (\ref{eq:3.2}) and (\ref{eq:4.17}), the proof of Theor.4.3 can be repeated obtaining the following limit, written in terms of overlaps thanks to eq. (\ref{eq:5.13}),
\begin{equation}\label{eq:5.19}
\lim_{x_{0}\rightarrow\infty}x_{0}e^{k_{n}x_{0}}\left(\varphi_{n},a_{\mathbf{x}_{0}}\psi_{0}\right)=d_{n}\left(\widehat{\mathbf{x}}_{0}\right)=-\frac{1}{4\pi}\int e^{k_{n}\widehat{\mathbf{x}}_{0}\cdot\mathbf{x}'_{0}}V_{\textrm{ext}}\left(\mathbf{x}'_{0}\right)\left(\varphi_{n},a_{\mathbf{x'}_{0}}\psi_{0}\right)d\mathbf{x}'_{0}.
\end{equation}
From the eqs. (\ref{eq:5.14}) and (\ref{eq:5.19}) it follows directly 
\begin{equation}\label{eq:5.20}
s-\lim_{x_{0}\rightarrow\infty}x_{0}e^{k_{n}x_{0}}\xi_{n,\mathbf{x}_{0}}=d_{n}\left(\widehat{\mathbf{x}}_{0}\right)\varphi_{n}
\end{equation}
which implies 
\begin{equation}\label{eq:5.21}
\lim_{x_{0}\rightarrow\infty}x_{0}^{2}e^{2k_{n}x_{0}}\rho_{n}\left(\mathbf{x}_{0}\right)=|d_{n}\left(\widehat{\mathbf{x}}_{0}\right)|^{2}.
\end{equation}

The limits of the eqs. (\ref{eq:5.19})-(\ref{eq:5.21}) are identical to the ones of the eqs. (\ref{eq:5.2}), (\ref{eq:5.5}) and (\ref{eq:5.7}) except for two remarkable differences.

($i$) They are subject to no energy cut-off related to the rate of the exponential decrease of the external potential. The reason is due to the mathematical structure of eq. (\ref{eq:5.16}), different from that of eq. (\ref{eq:3.2}) since the scalar product 
$(\varphi_{n},U^{(I)}_{\mathbf{r}} \psi_{0, \mathbf{r}})$ is replaced by the factorized term 
$V_{\textrm{ext}}\left(\mathbf{x}_{0}\right)\left(\varphi_{n},a_{\mathbf{x}_{0}}\psi_{0}\right)$.

($ii$) Even in the uncorrelated models bound states $\varphi_{n}$ embedded in the continuum are possible for particular combinations of values of the energies $e_{n}$. Differently from the correlated case they must belong necessarily to $P \mathscr{H}$ or to 
$Q \mathscr{H}$. This is purely due to the fact that the scattering states do not contribute to $\xi_{n, \mathbf{x}_{0}}$ and $\rho_{n}(\mathbf{x}_{0})$.

\subsection{Peculiarities of the subspace $P \mathscr{H}$ in the correlated systems}
According to the current opinion the property $\varphi_{n} \in P \mathscr{H}$ should be a peculiarity of the independent particle models and could not be shared by the correlated systems. This believe, contradicted by the mathematical proofs of this paper, is founded on the idea of a subspace $P \mathscr{H}$ structured as in the uncorrelated models. In this subsection we intend to show that in the correlated systems the structure of $P \mathscr{H}$ is quite different.

First let us examine the peculiarities of $P \mathscr{H}$ in the uncorrelated systems. As seen in Subsect. 5.5, $P \mathscr{H}$ has finite dimension $A+1$. Thus all its elements are single-hole states, i.e. they can be represented in the form $a_{f} \psi_{0}$ for some square summable function $f$. The existence in $P \mathscr{H}$ of states as $a_{\overline{\mathbf{x}}_{0}}\psi_{0}$ where 
$f\left(\mathbf{x}_{0}\right)=\delta\left(\mathbf{x}_{0}-\overline{\mathbf{x}}_{0}\right)\notin L^{2}$ must not be deceptive. Really, using eqs. (\ref{eq:5.12}) and (\ref{eq:5.13}) one can write also
\begin{equation}\label{eq:5.22}
a_{ \overline{\mathbf{x}}_{0}}\psi_{0}=\sum_{n=0}^{A}v_{n}\left(\overline{\mathbf{x}}_{0}\right)a_{v_{n}}\psi_{0}=a_{f_{n,\overline{\mathbf{x}}_{0}}}\psi_{0},
\qquad f_{n,\overline{\mathbf{x}}_{0}}( \mathbf{x}_{0})
=\sum_{n=0}^{A}\overline{v}_{n}\left( \mathbf{x}_{0}\right)v_{n}\left(\overline{\mathbf{x}}_{0}\right)\in L^{2}.
\end{equation}

In this subsection we study the subspace $P \mathscr{H}$ when the nucleons interact via a two-body potential obtaining the following results.

($i$) Both in Cartesian and Jacobi coordinates the dimension of $P \mathscr{H}$ is infinite.

($ii$) Due to its infinite dimension $P \mathscr{H}$ contains necessarily states which are strong limits of single-hole states but which cannot be represented in the form $a_{f} \psi_{0}$ for any $f \in L^{2}$. We call them \emph{generalized single-hole states} giving two necessary and sufficient conditions for their existence.

($iii$) The bound eigenstates of the residual nucleus are generalized single-hole states. This point is not supported by a theorem but by an argument which provides its evidence indirectly.

Let us mention in advance some mathematical properties which will be used below. The dimension of $P \mathscr{H}$ is given by the cardinality of the orthonormal set of the states $n_{\nu}^{- \frac{1}{2}} a_{\nu} \psi_{0}$ (with $n_{\nu} \neq 0$) which span this subspace, equal to the cardinality of the occupied natural orbitals $u_{\nu}$. We denote by $L^{2}_{0}(\mathbf{r})$ the subspace of 
$L^{2}(\mathbf{r})$ spanned by the occupied natural orbitals, and so having the same dimension as $P \mathscr{H}$. In view of the role played below by the density matrix $K$ let us recall here its general properties. It is always a nonnegative compact operator, defined in the whole $L^{2}(\mathbf{r})$, with finite trace $\sum_{\nu} n_{\nu} =A+1$. Although $K$ is not invertible in $L^{2}(\mathbf{r})$, its inverse exists when $K$ is restricted to $L^{2}_{0}(\mathbf{r})$. Due to general properties of the compact operators, the occupation number $n_{\nu}=0$ can have finite as well as infinite degeneracy, whereas the nonnull occupancies have always finite degeneracy. Thus if the occupied orbitals are infinite, also the related occupation numbers are infinite and must have $0$ as accumulation point due to the restriction $\sum_{\nu} n_{\nu} =A+1$. This makes critical the convergence of some series involving $n_{\nu}^{-1}$. Finally if the dimension of $L^{2}_{0}(\mathbf{r})$ is inf!
 inite the range $R(K)$ of $K$ is not a closed manifold. This point, essential in order that $P \mathscr{H}$ contains generalized single-hole states, is little known and requires a proof. 
\begin{thm}
If $L^{2}_{0}(\mathbf{r})$ has infinite dimension $R(K)$ is not closed.
\end{thm}
\begin{pft}
Let $\widehat{K}$ be the restriction of $K$ to $L^{2}_{0}(\mathbf{r})$. Since $R(K)$ coincides with $R( \widehat{K})$ it is sufficient to prove the theorem for $\widehat{K}$, working in the Hilbert space $L^{2}_{0}(\mathbf{r})$ where $\widehat{K} ^{-1}$ exists and is densely defined because its domain is $R( \widehat{K})$ which contains all the occupied orbitals. The proof is ab absurdo. Let us suppose that $R( \widehat{K})$ is closed which implies $R( \widehat{K})=L^{2}_{0}(\mathbf{r})$ so that $\widehat{K} ^{-1}$ is defined in the whole Hilbert space $L^{2}_{0}(\mathbf{r})$. As all the compact operators $\widehat{K}$ is closed, which implies that also 
$\widehat{K}^{-1}$ is closed. Therefore the theorem of the closed graph, applied to $\widehat{K}^{-1}$ which is closed and defined everywhere in $L^{2}_{0}(\mathbf{r})$, implies that $\widehat{K}^{-1}$ is a bounded operator. Thus, due to a general property of the compact operators, $\widehat{K} \widehat{K}^{-1}$ (which is the identity) is compact. This is absurd since the identity is not a compact operator.
\qedhere
\qedsymbol
\end{pft}

Of course if the number of the occupied orbitals is finite $L^{2}_{0}(\mathbf{r})$ has a finite dimension and the manifold 
$R( \widehat{K})$ is necessarily closed. Now, let us examine the points (i)-(iii).

\vspace{8pt}

($i$) \emph{Dimension of $P \mathscr{H}$}. We show that $P \mathscr{H}$ has infinite dimension proving that the occupied natural orbitals are infinite. To contain the number of the variables, we consider only the case of a three-body ground state $\psi_{0}$. For a more direct comparison with the independent particle model we use first Cartesian coordinates $\mathbf{x}_{0}$, $\mathbf{x}_{1}$, 
$\mathbf{x}_{2}$ fixing the center of mass motion by an additional external potential.

Let $v_{\nu}( \mathbf{x}_{0})$ and $n_{\nu}$ be the natural orbitals and occupation numbers of the density matrix
\begin{equation}\label{eq:5.23}
K\left(\mathbf{x}_{0},\mathbf{x}_{0}'\right)
=3\int\overline{\psi}_{0}\left(\mathbf{x}_{0}',\mathbf{x}_{1},\mathbf{x}_{2}\right)
\psi_{0}\left(\mathbf{x}_{0},\mathbf{x}_{1},\mathbf{x}_{2}\right)d\mathbf{x}_{1}d\mathbf{x}_{2}.
\end{equation}
The expansion of $\psi_{0}$ on the basis of the natural orbitals is
\begin{equation}\label{eq:5.24}
\psi_{0}\left(\mathbf{x}_{0},\mathbf{x}_{1},\mathbf{x}_{2}\right)=\sum_{\mu,\nu,\rho}c_{\mu,\nu,\rho}v_{\mu}\left(\mathbf{x}_{0}\right)v_{\nu}\left(\mathbf{x}_{1}\right)v_{\rho}\left(\mathbf{x}_{2}\right)
\end{equation}
with antisymmetric coefficients
\begin{equation}\label{eq:5.25}
c_{\mu,\nu,\rho}=\int\overline{v}_{\mu}\left(\mathbf{x}_{0}\right)\overline{v}_{\nu}\left(\mathbf{x}_{1}\right)\overline{v}_{\rho}\left(\mathbf{x}_{2}\right)\psi_{0}\left(\mathbf{x}_{0},\mathbf{x}_{1},\mathbf{x}_{2}\right)d\mathbf{x}_{0}d\mathbf{x}_{1}d\mathbf{x}_{2}.
\end{equation}
For $n_{\mu}=0$ one has $(v_{\mu},K \, v_{\mu})=0$ and hence, using eq. (\ref{eq:5.23}),
\begin{equation}\label{eq:5.26}
\int\overline{v}_{\mu}\left(\mathbf{x}_{0}\right)\psi_{0}\left(\mathbf{x}_{0},\mathbf{x}_{1},\mathbf{x}_{2}\right)d\mathbf{x}_{0}=0 \quad \forall \, 
\mathbf{x}_{1},\mathbf{x}_{2},
\end{equation}
which implies $c_{\mu,\nu,\rho}=0$ $\forall \nu, \rho$. This property is extended to the other indices by antisymmetry so that the expansion (\ref{eq:5.24}) contains only occupied orbitals. The antisymmetry of $c_{\mu,\nu,\rho}$  allows also to arrange the sum of eq. (\ref{eq:5.24}) in a weighted sum of Slater determinants of occupied orbitals which is finite if their number is finite and infinite otherwise. But in the presence of two-body interactions this sum is necessarily infinite because the eigenvalue equation for $\psi_{0}$ cannot have solutions which are finite linear combinations of factorized functions. This property is expressed in physical terms by the statement that the wave-functions of such a type represent partially correlated states. 

The case of three nucleons interacting only via a two-body interaction and so described by the Jacobi coordinates $\mathbf{r}$ and 
$\boldsymbol{\xi}$ requires a few changes. In order to have an expression of $\psi_{0}\left(\mathbf{r},\boldsymbol{\xi}\right)$ in terms of occupied natural orbitals one must use two different density matrices
\begin{equation}\label{eq:5.27}
K\left(\mathbf{r},\mathbf{r}'\right)=3\int\overline{\psi}_{0}\left(\mathbf{r}',\boldsymbol{\xi}\right)\psi_{0}\left(\mathbf{r},\boldsymbol{\xi}\right)d\boldsymbol{\xi},\qquad 
K'\left(\boldsymbol{\xi},\boldsymbol{\xi}'\right)=3\int\overline{\psi}_{0}\left(\mathbf{r},\boldsymbol{\xi}'\right)\psi_{0}\left(\mathbf{r},\boldsymbol{\xi}\right)d\mathbf{r},
\end{equation}
respectively with natural orbitals $u_{\mu}$, $u_{\nu}'$ and occupancies $n_{\mu}$, $n_{\nu}'$. The first one is the density matrix used habitually in this paper. One has
\begin{equation}\label{eq:5.28}
\psi_{0}\left(\mathbf{r},\boldsymbol{\xi}\right)=\sum_{\mu,\nu}c_{\mu,\nu}u_{\mu}\left(\mathbf{r}\right)u_{\nu}'\left(\boldsymbol{\xi}\right)
\end{equation}
where the antisymmetry property, hidden into the variables, does not affect the coefficients. The expansion (\ref{eq:5.28}) contains only occupied orbitals since for $n_{\mu} = n_{\nu}' = 0$ the relations $(u_{\mu},K \, u_{\mu})=(u_{\nu}',K' \, u_{\nu}')=0$ yields
\begin{equation}\label{eq:5.29}
\int\overline{u}_{\mu}\left(\mathbf{r}\right)\psi_{0}\left(\mathbf{r},\boldsymbol{\xi}\right)d\mathbf{r}=0 \quad \forall \, \boldsymbol{\xi},
\qquad 
\int\overline{u}_{\nu}'\left(\boldsymbol{\xi}\right)\psi_{0}\left(\mathbf{r},\boldsymbol{\xi}\right)d\boldsymbol{\xi}=0 \quad \forall \, \mathbf{r}.
\end{equation}
The proof that both the occupied orbitals $u_{\mu}$ and $u_{\nu}'$ are infinite follows from the same argument as above.

\begin{conclusion}
Whether one use Cartesian or Jacobi coordinates, in the fully correlated systems the dimension of $P \mathscr{H}$ is infinite and the occupation numbers of the density matrices $K$ and $K'$ accumulate at $0$. Besides, since in both cases the occupied orbitals are infinite, the ranges of the density matrices $K$ and $K'$ are not closed by Theor.5.1.
\end{conclusion}

\vspace{8pt}

($ii$) \emph{Generalized single-hole states}. Here we use only occupied natural orbitals $u_{\nu}$, ordered by decreasing values of the occupancies $n_{\nu}$. Let us consider first an uncorrelated system where the number $N=A+1$ of the occupied 
$u_{\nu}$ is finite and equal to the dimension of 
$P \mathscr{H}$. Every state $\varphi$ of $P \mathscr{H}$ satisfies the relation $\varphi = P \varphi$ and by eq. (\ref{eq:2.9}) can be expressed in the form
\begin{equation}\label{eq:5.30}
\varphi=\sum_{\nu=1}^{N}\left(a_{\nu}\psi_{0},\varphi\right)\frac{1}{n_{\nu}}a_{\nu}\psi_{0}=a_{f_{N}}\psi_{0},
\end{equation}
where
\begin{equation}\label{eq:5.31}
f_{N}\left(\mathbf{r}\right)
=\sum_{\nu=1}^{N}\left(\varphi, a_{\nu}\psi_{0}\right)\frac{1}{n_{\nu}}u_{\nu}\left(\mathbf{r}\right)\in L^{2}\left(\mathbf{r}\right).
\end{equation}
Thus $P \mathscr{H}$ is composed exclusively of single-hole states. Moreover, since 
$\left(a_{f_{N}}\psi_{0},a_{\mathbf{r}}\psi_{0}\right)$
$=\left(K \, f_{N}\right)\left(\mathbf{r}\right)$,
every overlap $\left(\varphi,a_{\mathbf{r}}\psi_{0}\right)$ belongs to the range $R(K)$ of $K$. Now, let us consider the case of the correlated systems where $N= \infty$ and every state $\varphi \in P \mathscr{H}$ is expressed by the strong limit of single-hole states
\begin{equation}\label{eq:5.32}
\varphi=\sum_{\nu=1}^{\infty}\left(a_{\nu}\psi_{0},\varphi\right)\frac{1}{n_{\nu}}a_{\nu}\psi_{0}=s-\lim_{N\rightarrow\infty}a_{f_{N}}\psi_{0},
\end{equation}
where $f_{N}$ is given still by eq. (\ref{eq:5.31}). Using eq. (\ref{eq:5.32}) and the Schwarz inequality one has, in the sense of the strong convergence in $L^{2}( \mathbf{r})$,
\begin{equation}\label{eq:5.33}
\left(\varphi,a_{\mathbf{r}}\psi_{0}\right)=s-\lim_{N\rightarrow\infty}\left(a_{f_{N}}\psi_{0},a_{\mathbf{r}}\psi_{0}\right)=s-\lim_{N\rightarrow\infty}\left(K \, f_{N}\right)\left(\mathbf{r}\right),\quad\forall\varphi\in P\mathscr{H}.
\end{equation}
This implies that $\left(\varphi,a_{\mathbf{r}}\psi_{0}\right)$ belongs to the closure $\overline{R(K)}$ of $R(K)$. Note that every occupied orbital satisfies the relation $u_{\nu}=n_{\nu}^{-1}K \, u_{\nu}$ and so belongs to $R(K)$. Therefore $\overline{R(K)}$ coincides with $L^{2}_{0}( \mathbf{r})$.

Now we estabilish two characteristic conditions in order that a vector $\varphi \in P \mathscr{H}$ is a generalized single-hole state, i.e. a state which cannot be represented in the form $a_{f} \psi_{0}$ for any $f \in L^{2}$.

\begin{condition}
A state $\varphi \in P \mathscr{H}$ is a generalized single-hole state if and only if $(\varphi , a_{\mathbf{r}} \psi_{0})$ does not belong to the range $R(K)$ of the density matrix $K$.
\end{condition}
\begin{pft}
The proof is ab absurdo. Let $\varphi$ be a generalized single-hole state. If $(\varphi , a_{\mathbf{r}} \psi_{0})$ belongs to $R(K)$ there exists a function $f \in L^{2}( \mathbf{r})$ such that $(\varphi , a_{\mathbf{r}} \psi_{0})=(K \, f)( \mathbf{r})$. This implies
$(a_{\nu} \psi_{0}, \varphi)=n_{\nu}(f, u_{\nu})$ which, inserted in the first eq. (\ref{eq:5.32}), yields $\varphi = a_{f} \psi_{0}$ and contradicts the hypothesis. Conversely, let us suppose $(\varphi , a_{\mathbf{r}} \psi_{0}) \notin R(K)$. If $\varphi$ is a single-hole state there exists a function $f \in L^{2}( \mathbf{r})$ such that $\varphi = a_{f} \psi_{0}$ and one has 
$(\varphi , a_{\mathbf{r}} \psi_{0})$ $=( a_{f} \psi_{0} , a_{\mathbf{r}} \psi_{0})$ $=(K \, f)( \mathbf{r})$. Thus 
$(\varphi , a_{\mathbf{r}} \psi_{0})$ belongs to $R(K)$ which contradicts the hypothesis.
\qedhere
\qedsymbol
\end{pft}

The condition 1 shows that a generalized single-hole state $\varphi$ is characterized by the conditions 
$(\varphi , a_{\mathbf{r}} \psi_{0}) \notin R(K)$ and $(\varphi , a_{\mathbf{r}} \psi_{0}) \in \overline{R(K)}=L^{2}_{0}( \mathbf{r})$. This is possible only if the dimension of $P \mathscr{H}$ is infinite so that, due to Theor.5.1, $R(K)$ is not closed and so is contained strictly in $\overline{R(K)}$. The real existence of generalized single-hole states is proved by a second characteristic condition.
\begin{condition}
A state $\varphi \in P \mathscr{H}$ is a generalized single-hole state if and only if
\begin{equation}\label{eq:5.34}
\sum_{\nu=1}^{\infty}
\Big|\frac{\left( \varphi, a_{\nu}\psi_{0} \right)}{n_{\nu}}\Big|^{2}=\infty.
\end{equation}
\end{condition}
\begin{pft}
The proof is ab absurdo. Let $\varphi$ be a generalized single-hole state. If the series of eq. (\ref{eq:5.34}) converges, the functions $f_{N}$ of eq. (\ref{eq:5.31}) converge strongly in $L^{2}( \mathbf{r})$ to a function $f$ of $L^{2}( \mathbf{r})$. Using the Schwarz inequality one has
\begin{equation}\label{eq:5.35}
\lim_{N\rightarrow\infty}||a_{f}\psi_{0}-a_{f_{N}}\psi_{0}||^{2}\leq\int||a_{\mathbf{r}}\psi_{0}||^{2}d\mathbf{r}\lim_{N\rightarrow\infty}\int|f\left(\mathbf{r}\right)-f_{N}\left(\mathbf{r}\right)|^{2} d\mathbf{r}=0
\end{equation} 
so that $a_{f_{N}}\psi_{0}$ converges strongly to $a_{f} \psi_{0}$. Thus by eq. (\ref{eq:5.32}) it is $\varphi = a_{f} \psi_{0}$ which contradicts the hypothesis. Conversely, let us suppose that eq. (\ref{eq:5.34}) holds. If $\varphi$ is a single-hole state there exists a function $f \in L^{2}( \mathbf{r})$ such that $\varphi = a_{f} \psi_{0}$. Therefore one has 
$( \varphi, a_{\nu} \psi_{0})= n_{\nu}(u_{\nu},f)$ so that the series of eq. (\ref{eq:5.34}) becomes 
$\sum_{\nu =1}^{\infty}|(u_{\nu},f)|^{2}$ which converges and contradicts the hypothesis.
\qedhere
\qedsymbol
\end{pft}

The condition 2 is the tool to prove that $P \mathscr{H}$ contains a lot of generalized single-hole states. For instance let us consider the states
\begin{equation}\label{eq:5.36}
\varphi=\sum_{\nu=1}^{\infty}c_{\nu}a_{\nu}\psi_{0}\quad\textrm{with}\quad\sum_{\nu=1}^{\infty}n_{\nu}|c_{\nu}|^{2}<\infty\quad\textrm{and}\quad\sum_{\nu=1}^{\infty}|c_{\nu}|^{2}=\infty,
\end{equation} 
where the convergence of the second series implies the convergence of the first one since the vectors 
$n_{\nu}^{- \frac{1}{2}} a_{\nu} \psi_{0}$ form an orthonormal system. All these states belong to $P \mathscr{H}$ and are generalized single-hole states due to the divergence of the third series which implies eq. (\ref{eq:5.34}) since 
\begin{equation}\label{eq:5.37}
\sum_{\nu=1}^{\infty}
\Big|\frac{\left( \varphi,a_{\nu}\psi_{0} \right)}{n_{\nu}}\Big|^{2}
=\sum_{\nu=1}^{\infty}|c_{\nu}|^{2}=\infty.
\end{equation} 
The states $a_{\overline{\mathbf{r}}} \psi_{0}$ themselves, if nonnull, are generalized single-hole states because 
$|n_{\nu}^{-1}\left(a_{\overline{\mathbf{r}}}\psi_{0},a_{\nu}\psi_{0}\right)|^{2}=|u_{\nu}\left(\overline{\mathbf{r}}\right)|^{2}$
and the series $\sum_{\nu=1}^{\infty}|u_{\nu}\left(\overline{\mathbf{r}}\right)|^{2}$ does not converge. This difference from the independent particle model, where the states $a_{\overline{\mathbf{r}}} \psi_{0}$ are single-hole states as stressed by eq. 
(\ref{eq:5.22}), shows that the distinction between the single-hole states and the generalized ones may depend only on the dimension of 
$P \mathscr{H}$. Note that many other states $a_{f} \psi_{0}$ with $f \notin L^{2}( \mathbf{r})$ share this peculiarity, e.g. if $f$ is a plane wave or grows at the infinity as $e^{\alpha r}$ with $\alpha < k_{0}$.

\vspace{8pt}

($iii$) \emph{Bound eigenstates of H belonging to $P \mathscr{H}$}. We have not a direct proof, based on the condition 1 or 2, that the eigenstates $\varphi_{n}$ are generalized single-hole states. Thus we assume the opposite thesis that 
$\varphi_{n}=a_{f_{n}} \psi_{0}$ for a $f_{n} \in L^{2}$, find an equation that should be satisfied by $f_{n}$ and discuss whether it can have an acceptable solution. To have a direct comparison with the uncorrelated case we disregard the center of mass motion and operate with Cartesian coordinates. To extend easily the present treatment to the case of Jacobi coordinates we restrict the role of the Fock operators to the relations
\begin{equation}\label{eq:5.38}
\left(a_{\mathbf{x}_{0}}\psi_{0}\right)\left(\mathbf{x}\right)=\sqrt{A+1}\psi_{0}\left(\mathbf{x}_{0},\mathbf{x}\right),\quad\left(a_{\mathbf{x}_{0}}a_{\mathbf{x}_{1}}\psi_{0}\right)\left(\mathbf{y}\right)=\sqrt{A\left(A+1\right)}\psi_{0}\left(\mathbf{x}_{0},\mathbf{x}_{1},\mathbf{y}\right),
\end{equation} 
where $\mathbf{x}$ and $\mathbf{y}$ represent the residual variables, and we do not use the canonical anticommutation rules.

If $\varphi_{n}=a_{f_{n}} \psi_{0}$, for a $f_{n} \in L^{2}$, one has
\begin{equation}\label{eq:5.39}
\int\left(Ha_{\mathbf{x}_{0}'}\psi_{0},a_{\mathbf{x}_{0}}\psi_{0}\right)f_{n}\left(\mathbf{x}_{0}'\right)d\mathbf{x}_{0}'=\varepsilon_{n}\left(a_{f_{n}}\psi_{0},a_{\mathbf{x}_{0}}\psi_{0}\right)=\varepsilon_{n}\left(Kf_{n}\right)\left(\mathbf{x}_{0}\right).
\end{equation} 
Let us decompose the $(A+1)$-body Hamiltonian $H^{(A+1)}$ into $H$, the kinetic energy of the nucleon and the interaction 
$V_{\mathbf{x}_{0}}^{(I)}=\sum_{i=1}^{A}V_{\mathbf{x}_{0}}^{(i)}$
with $V_{\mathbf{x}_{0}}^{(i)}\left(\mathbf{x}_{i}\right)=V\left(\mathbf{x}_{0}-\mathbf{x}_{i}\right)$. Therefore eq. 
(\ref{eq:5.39}) reads
\begin{equation}\label{eq:5.40}
\frac{1}{2m}\int\left(\boldsymbol{\nabla}_{\mathbf{x}_{0}'}^{2}K\left(\mathbf{x}_{0},\mathbf{x}_{0}'\right)\right)f_{n}\left(\mathbf{x}_{0}'\right)d\mathbf{x}_{0}'=\left(\varepsilon_{n}-E_{0}\right)\left(Kf_{n}\right)\left(\mathbf{x}_{0}\right)+\int \mathcal{V}^{(I)}\left(\mathbf{x}_{0},\mathbf{x}_{0}'\right)f_{n}\left(\mathbf{x}_{0}'\right)d\mathbf{x}_{0}',
\end{equation}  
where
\begin{equation}\label{eq:5.41}
\mathcal{V}^{(I)}\left(\mathbf{x}_{0},\mathbf{x}_{0}'\right)\equiv\left(V_{\mathbf{x}_{0}'}^{(I)}a_{\mathbf{x}_{0}'}\psi_{0},a_{\mathbf{x}_{0}}\psi_{0}\right).
\end{equation}  
Since every component of $V_{\mathbf{x}_{0}'}^{(I)}$ yields the same contribution as $V_{\mathbf{x}_{0}'}^{(1)}$ one has, by the pure use of the eqs. (\ref{eq:5.38}),
\begin{alignat}{1}\label{eq:5.42}
\mathcal{V}^{(I)}\left(\mathbf{x}_{0},\mathbf{x}_{0}'\right) & =A\left(V_{\mathbf{x}_{0}'}^{(1)}a_{\mathbf{x}_{0}'}\psi_{0},a_{\mathbf{x}_{0}}\psi_{0}\right)=
\int \left(V_{\mathbf{x}_{0}'}^{(1)}a_{\mathbf{x}_{0}'}a_{\mathbf{x}_{1}}\psi_{0},a_{\mathbf{x}_{0}}a_{\mathbf{x}_{1}}\psi_{0}\right) d\mathbf{x}_{1} \nonumber \\
 & =\int V\left(\mathbf{x}_{0}'-\mathbf{x}_{1}\right)\left(a_{\mathbf{x}_{0}'}a_{\mathbf{x}_{1}}\psi_{0},a_{\mathbf{x}_{0}}a_{\mathbf{x}_{1}}\psi_{0}\right)d\mathbf{x}_{1}.  
\end{alignat}
Assuming that one may integrate by parts in the left-hand side of eq. (\ref{eq:5.40}), it follows that $f_{n}$ must satisfy the equation
\begin{equation}\label{eq:5.43}
\frac{1}{2m}K\boldsymbol{\nabla}^{2}f_{n}=\left(\varepsilon_{n}-E_{0}\right)Kf_{n}+\mathcal{V}^{\left(I\right)}f_{n}
\end{equation}  
which can have a solution in $L^{2}( \mathbf{r})$ only if $\mathcal{V}^{\left(I\right)}f_{n}$ belongs to the range $R(K)$ of the density matrix.

Eq. (\ref{eq:5.43}) holds also in the uncorrelated models where $V_{\mathbf{x_{0}}}^{(I)}$ coincides with the external potential 
$V_{\textrm{ext}}\left(\mathbf{x}_{0}\right)$ and $\mathcal{V}^{(I)}\left(\mathbf{x}_{0},\mathbf{x}_{0}'\right)$
with $K\left(\mathbf{x}_{0},\mathbf{x}_{0}'\right)V_{\textrm{ext}}\left(\mathbf{x}_{0}'\right)$. Thus one has 
$\mathcal{V}^{(I)}f_{n}=KV_{\textrm{ext}}f_{n}$ which belongs to $R(K)$, as necessary. Hence, applying the operator $K^{-1}$ to both sides of eq. (\ref{eq:5.43}), one recovers the well-known result that $f_{n}$ is an eigenfunction of the one-body Hamiltonian for the external potential.

In the case of two-body interactions it is necessary to check whether $\mathcal{V}^{\left(I\right)}f_{n}$ belongs to $R(K)$. This requires that in the basis of the natural orbitals it is
\begin{equation}\label{eq:5.44}
\sum_{\nu=1}^{\infty}\frac{|\left(u_{\nu},\mathcal{V}^{(I)}f_{n} \right)|^{2}}{n_{\nu}^{2}}<\infty.
\end{equation}  
Due to eq. (\ref{eq:5.42}) one has
\begin{equation}\label{eq:5.45}
\left(u_{\nu},\mathcal{V}^{(I)}f_{n} \right)=\int\left(a_{\mathbf{x}_{0}'}a_{\mathbf{x}_{1}}\psi_{0},a_{\nu}a_{\mathbf{x}_{1}}\psi_{0}\right)V\left(\mathbf{x}_{0}'-\mathbf{x}_{1}\right)f_{n} \left(\mathbf{x}_{0}'\right)d\mathbf{x}_{0}'
d\mathbf{x}_{1}
\end{equation}  
which can be expressed as the A-body scalar product $(\phi_{n} , a_{\nu} \psi_{0})$ where
\begin{equation}\label{eq:5.46}
\phi_{n} =\int V\left(\mathbf{x}_{0}'-\mathbf{x}_{1}\right)f_{n} \left(\mathbf{x}_{0}'\right)a_{\mathbf{x}_{1}}^{\dagger}a_{\mathbf{x}_{1}}a_{\mathbf{x}_{0}'}\psi_{0}d\mathbf{x}_{0}'d\mathbf{x}_{1}.
\end{equation}  
This implies the convergence of the series
\begin{equation}\label{eq:5.47}
\sum_{\nu=1}^{\infty}\frac{|\left(u_{\nu},\mathcal{V}^{(I)}f_{n} \right)|^{2}}{n_{\nu}}=\sum_{\nu=1}^{\infty}\frac{|\left(\phi_{n},a_{\nu}\psi_{0}\right)|^{2}}{n_{\nu}}=\left(\phi_{n},P\phi_{n} \right),
\end{equation}  
but one can find no reason to justify the stronger condition (\ref{eq:5.44}) on the basis of eq. (\ref{eq:5.46}). In conclusion one cannot prove that eq. (\ref{eq:5.43}) has a solution in $L^{2}( \mathbf{x}_{0})$. On the contrary there is a significant indication that this is not possible so that the bound states $\varphi_{n}$ belonging to $P \mathscr{H}$  are generalized single-hole states. In such a case the real difference between the uncorrelated and correlated systems is that in the latter ones the picture of $\varphi_{n}$ as a hole in $\psi_{0}$ is not appropriate since one cannot assign a wave funtion to the hole. This distinction is conceptually important and sufficient to answer the objection that the bound states cannot have identical properties in the uncorrelated and correlated systems.

\begin{nota}
The adjoint operator $a_{\mathbf{x}_{1}}^{\dagger}$ in eq. (\ref{eq:5.46}) can be defined also in Jacobi coordinates. The essential difference is that $a_{\mathbf{x}}^{\dagger}$ and $a_{\mathbf{y}}$ do not satisfy the canonical anticommutation rules. Since these rules have not been used, the present conclusions can be extended to the case of Jacobi coordinates.
\end{nota}

\section{Projection operators for degenerate eigen\-values}

In this section we consider spherically symmetric two-body potentials disregarding the spin depending interactions so that $\psi_{0}$ represents the orbital factor of the $(A+1)$-body ground-state. If the O'Connor-Simon upper bound holds also in presence of such interactions, as expected but at present lacking a rigorous proof, it is sufficient to replace the partial-wave expansion used here by the angular-spin decomposition. The center of mass motion, neglected in \cite{VWH} where the V.W.H. method was presented first, is removed here by means of Jacobi coordinates. The changes necessary to operate in Cartesian coordinates are direct. This section has a twofold motivation: ($i$) To give a rigorous mathematical justification to the V.W.H. method for discrete energies of the residual nucleus lower than $\min\left\{ \varepsilon_{C},\varepsilon_{M}\right\} $. ($ii$) To treat a concrete case of degeneracy of the eigenvalues. Really, if the eigenvalue is degene\-rate, the treatme!
 nt of Sects. 4 and 5 provides a single unspecified eigenvector of the degenerate subspace (which can  depend on the direction $\widehat{ \mathbf{r}}$ of the limit). Operating with partial waves one can remedy this drawback and introduce more specialized projection operators onto subspaces of 
$P \mathscr{H}$ which define better the nature of the bound states of the residual nucleus.

Let us consider the expansion on the basis of the spherical harmonics $Y^{(l,m)}$
\begin{equation}\label{eq:6.1}
a_{\mathbf{r}}\psi_{0}=\sum_{l=0}^{\infty}\sum_{m=-l}^{l}a_{r}^{\left(l,m\right)}\psi_{0}Y^{\left(l,m\right)}\left(\widehat{\mathbf{r}}\right),\quad a_{r}^{\left(l,m\right)}\psi_{0}=\int\overline{Y^{\left(l,m\right)}}\left(\widehat{\mathbf{r}}\right)a_{\mathbf{r}}\psi_{0}d\Omega_{\widehat{\mathbf{r}}},
\end{equation}  
where  $a_{r}^{\left(l,m\right)}\psi_{0}$ has the same meaning of mere notation as $a_{ \mathbf{r}}\psi_{0}$. Note that in Cartesian coordinates $a_{ \mathbf{x}_{0}}^{\left(l,m\right)}$ is the annihilation operator divided by $x_{0}$. Since the potential is spherically symmetric and $\psi_{0}$ is not degenerate (see below eq. (\ref{eq:2.6})), it has angular momentum zero. Therefore the bound eigenstates of the residual nucleus can be characterized by spherical single-particle quantum numbers $l$, $m$ and be denoted by $\varphi_{n}^{(l,m)}$. Due to the first  eq. (\ref{eq:2.11}) the overlaps $(\varphi_{n}^{(l,m)},a_{ \mathbf{r}}\psi_{0})$ are eigenfunctions of the energy-dependent hole Hamiltonian $H^{(h)}$, which is rotationally invariant. Thus they are factorized as 
\begin{equation}\label{eq:6.2}
\left(\varphi_{n}^{\left(l,m\right)},a_{\mathbf{r}}\psi_{0}\right)=\phi_{n}^{\left(l\right)}\left(r\right)Y^{\left(l,m\right)}\left(\widehat{\mathbf{r}}\right),\quad\phi_{n}^{\left(l\right)}\left(r\right)\equiv\left(\varphi_{n}^{\left(l,m\right)},a_{r}^{\left(l,m\right)}\psi_{0}\right),
\end{equation}  
where $\phi_{n}^{\left(l\right)}\left(r\right)$ is independent on $m$ as well as the energy $\varepsilon^{(l)}_{n}$ of 
$\varphi_{n}^{(l,m)}$. The rotational invariance of the density matrix 
$K( \mathbf{r}, \mathbf{r}')=(a_{ \mathbf{r}'}\psi_{0},a_{ \mathbf{r}}\psi_{0})$ implies
\begin{equation}\label{eq:6.3}
\left(a_{r'}^{\left(l',m'\right)}\psi_{0},a_{r}^{\left(l,m\right)}\psi_{0}\right)=\delta_{l,l'}\delta_{m,m'}K_{0}^{\left(l\right)}\left(r,r'\right)
\end{equation}  
where the partial-wave density matrix $K_{0}^{\left(l\right)}\left(r,r'\right)$ is independent on $m$. We use also the subtracted density matrices
\begin{equation}\label{eq:6.4}
K_{n}^{\left(l\right)}\left(r,r'\right)=K_{0}^{\left(l\right)}\left(r,r'\right)-\sum_{n'=0}^{n-1}\overline{\phi_{n'}^{\left(l\right)}}\left(r'\right)\phi_{n'}^{\left(l\right)}\left(r\right).
\end{equation}  
Let us introduce the partial-wave components of the vector $\xi_{n, \mathbf{r}}$ defined in eq. (\ref{eq:5.4}):
\begin{equation}\label{eq:6.5}
\xi_{n,r}^{\left(l,m\right)}=\int\overline{Y^{\left(l,m\right)}}\left(\widehat{\mathbf{r}}\right)\xi_{n,\mathbf{r}}d\Omega_{\widehat{\mathbf{r}}}=a_{r}^{\left(l,m\right)}\psi_{0}-\sum_{n'=0}^{n-1}\phi_{n'}^{\left(l\right)}\left(r\right)\varphi_{n'}^{\left(l,m\right)}.
\end{equation}

The asymptotic behaviours of $\phi_{n}^{\left(l\right)}$, $\xi_{n,r}^{\left(l,m\right)}$
and $K_{n}^{\left(l\right)}$ are deduced easily from results
of Sect. 4. Let us set $k_{n}^{\left(l\right)}=\left(2\mu_{0}\left(\varepsilon_{n}^{(l)}-E_{0}\right)\right)^{\frac{1}{2}}$
. From eq. (\ref{eq:6.2}) written for $m=0$ one has
\begin{equation}\label{eq:6.6}
r\, e^{k_{n}^{\left(l\right)}r}\phi_{n}^{\left(l\right)}\left(r\right)=\int r\, e^{k_{n}^{\left(l\right)}r}\left(\varphi_{n}^{\left(l,0\right)},a_{\mathbf{r}}\psi_{0}\right)\overline{Y^{\left(l,0\right)}}\left(\widehat{\mathbf{r}}\right)d\Omega_{\widehat{\mathbf{r}}}
\end{equation}  
where using eq. (\ref{eq:4.29}) the absolute value of the integrand is majored by a multiple of $|Y^{(l,0)}( \widehat{\mathbf{r}})|$, independent on $r$ and summable over the angles. Therefore by dominated convergence one exchanges the limit for $r \rightarrow\infty$ and the integral of eq. (\ref{eq:6.6}) obtaining by eq. (\ref{eq:5.2}) (with $\lambda$ replaced by $n$, $l$, 0)
\begin{equation}\label{eq:6.7}
\lim_{r\rightarrow\infty}r\, e^{k_{n}^{\left(l\right)}r}\phi_{n}^{\left(l\right)}\left(r\right)=
d_{n}^{\left(l\right)}\equiv
\int\overline{Y^{\left(l,0\right)}}\left(\widehat{\mathbf{r}}\right)d^{(l,0)}_{n}\left(\widehat{\mathbf{r}}\right)
d\Omega_{\widehat{\mathbf{r}}},\;\textrm{for}\;\varepsilon_{n}^{(l)}<\varepsilon_{M}.
\end{equation}  
In eq. (\ref{eq:6.7}) the dependence of $d^{(l,0)}_{n}\left(\widehat{\mathbf{r}}\right)$ on $\widehat{\mathbf{r}}$ is crucial: without it only $d_{n}^{(0)}$ would be different from zero (since the integrals of the Legendre polynomials $Y^{(l,0)}$ with $l \neq 0$ are zero) so that the V.W.H. method would give results only for bound states with null angular momentum. The asymptotic behaviour of 
$\xi_{n,r}^{(l,m)}$ follows readily from the first eq. (\ref{eq:6.5}) and eq. (\ref{eq:4.35}) which yield, for every 
$\varphi \in \mathscr{H}$,
\begin{equation}\label{eq:6.8}
|\left(\varphi,\xi_{n+1,r}^{\left(l,m\right)}\right)|
\leq \sqrt{A+1} \, B_{k,q}'C^{\left(l,m\right)}||\varphi||e^{-qr},
\quad\forall q<\min\left\{ k_{n+1}^{\left(l\right)},k_{C},k_{M}\right\} ,
\end{equation}   
where $C^{\left(l,m\right)}$ is the integral of $|Y^{\left(l,m\right)}( \widehat{ \mathbf{r}})|$ over the angles and the restriction $q < k^{(l)}_{n+1}$ takes into account that $a^{(l,m)}_{r} \psi_{0}$ does not contain the contribution of the bound states 
$\varphi^{(l',m')}_{n'}$ with $l' \neq l$ and $m' \neq m$ due to eq. (\ref{eq:6.2}). Therefore the same steps we made below eq. (\ref{eq:4.35}) yield 
\begin{equation}\label{eq:6.9}
s-\lim_{r\rightarrow\infty}
r\, e^{k_{n}^{\left(l\right)}r}\xi_{n,r}^{\left(l,m\right)}=
d_{n}^{\left(l\right)}\varphi_{n}^{\left(l,m\right)},
\quad\forall\varepsilon_{n}^{(l)}<\min\left\{ \varepsilon_{C},\varepsilon_{M}\right\} ,
\end{equation}   
which implies the corresponding weak convergence and the one of the related norms. Since it is
\begin{equation}\label{eq:6.10}
K_{n}^{\left(l\right)}\left(r,r'\right)=\left(a_{r'}^{\left(l,m\right)}\psi_{0},\xi_{n,r}^{\left(l,m\right)}\right),\quad\rho_{n}^{\left(l\right)}\left(r\right)\equiv K_{n}^{\left(l\right)}\left(r,r\right)=
||\xi_{n,r}^{\left(l,m\right)}||^{2},
\end{equation}   
one obtains automatically the asymptotic behaviour of the (subtracted) density matrices and nuclear densities:
\begin{equation}\label{eq:6.11}
\lim_{r\rightarrow\infty}r\, e^{k_{n}^{\left(l\right)}r}K_{n}^{\left(l\right)}\left(r,r'\right)=
d_{n}^{\left(l\right)} \overline{ \phi}_{n}^{\left(l\right)}\left(r' \right),
\quad\lim_{r\rightarrow\infty}r^{2}e^{2k_{n}^{\left(l\right)}r}\rho_{n}^{\left(l\right)}\left(r\right)=|d_{n}^{\left(l\right)}|^{2}.
\end{equation}   
The eqs. (\ref{eq:6.11}), proved here for $\varepsilon_{n}^{(l)}<\min\left\{ \varepsilon_{C},\varepsilon_{M}\right\} $, are the basic assumptions of the V.W.H. method which can work only if $d_{n}^{(l)}\neq 0 $. To our knowledge in the calculations based on this method there is no gap of eigenvalues and related overlaps. This suggests indirectly that the anomalies $d_{n}^{(l)} = 0 $ should be rare events in Nuclear Physics, where also the restriction $\varepsilon_{n} < \varepsilon_{M}$ is not severe, as observed in Subsect. 5.1. Note that the exclusion of the bound states embedded in the continuum of $H$ is an absolute restriction. Sometimes the V.W.H. method is applied also to these states. This should be considered as an approximation in which the contribution of the scattering states below their energy is neglected. 

Let us compare the treatment of Sect. 5 and the present one in the hypothesis $d_{n}^{(l)} \neq 0$, so that the coefficients 
$d_{n}^{(l,0)} \left(\widehat{\mathbf{r}}\right)$ are not identically null. Both the treatments are subject to the same restriction 
$\varepsilon_{n}^{(l)}<\min\left\{ \varepsilon_{C},\varepsilon_{M}\right\} $. For every direction $\widehat{\mathbf{r}}$, eq. (\ref{eq:5.5}) yields an unique eigenvector of the degenerate eigenvalue $\varepsilon_{n}^{(l)}$ and establishes that it belongs to the subspace $P \mathscr{H}$. Changing the direction $\widehat{\mathbf{r}}$ of the limit one obtains various eigenvectors but no information about them and their relations. On the contrary eq. (\ref{eq:6.9}) provides all the eigenvectors $\varphi_{n}^{(l,m)}$ and allows to distinguish the subspaces where they live by means of the more specialized projection operators
\begin{equation}\label{eq:6.12}
P^{\left(l,m\right)}\varphi=\sum_{\nu,\; n_{\nu}^{(l)}\neq0}\left(a_{\nu}^{\left(l,m\right)}\psi_{0},\varphi\right)\frac{1}{n_{\nu}^{(l)}}a_{\nu}^{\left(l,m\right)}\psi_{0},\quad a_{\nu}^{\left(l,m\right)}\psi_{0}\equiv\int_{0}^{\infty}\overline{u}_{\nu}^{(l)}\left(r\right)a_{r}^{\left(l,m\right)}\psi_{0}dr,
\end{equation}   
where $u_{\nu}^{(l)}$ and $n_{\nu}^{(l)}$ are the natural orbitals and the related occupancies of the density matrix $K_{0}^{(l)}$. The proof that the eigenvectors $\varphi_{n}^{(l,m)}$ belong to the subspace $P^{(l,m)} \mathscr{H}$ is the same as in Sect. 5. Due to eq. (\ref{eq:6.3}) the projection operators $P^{(l,m)}$ are mutually orthogonal so that
\begin{equation}\label{eq:6.13}
P^{\left(l,m\right)}a_{r}^{\left(l',m'\right)}\psi_{0}=\delta_{l,l'}\delta_{m,m'}a_{r}^{\left(l,m\right)}\psi_{0}.
\end{equation} 
Besides $P \mathscr{H}$ is the direct sum of the subspaces $P^{(l,m)} \mathscr{H}$: 
\begin{equation}\label{eq:6.14}
P\mathscr{H}=\sum_{l=0}^{\infty}\sum_{m=-l}^{l}\oplus P^{\left(l,m\right)}\mathscr{H}.
\end{equation} 
In fact using the first eq. (\ref{eq:6.1}) and eq. (\ref{eq:6.13}) one has for every $\mathbf{r}$
\begin{alignat}{1}\label{eq:6.15}
\sum_{l,m}P^{\left(l,m\right)}a_{\mathbf{r}}\psi_{0} & 
=\sum_{l,m}\sum_{l',m'}P^{\left(l,m\right)}a_{r}^{\left(l',m'\right)}\psi_{0}
Y^{\left(l',m' \right)}\left(\widehat{\mathbf{r}}\right) \nonumber \\
 & =\sum_{l,m}a_{r}^{\left(l,m\right)}\psi_{0}Y^{\left(l,m\right)}\left(\widehat{\mathbf{r}}\right)=a_{\mathbf{r}}\psi_{0}=Pa_{\mathbf{r}}\psi_{0}. 
\end{alignat}
Since $P \mathscr{H}$ is spanned by the vectors $a_{ \mathbf{r}} \psi_{0}$, eq. (\ref{eq:6.15}) holds also when the operators are applied to an arbitrary $\varphi \in P \mathscr{H}$ and eq. (\ref{eq:6.14}) follows.

Using the eqs. (\ref{eq:6.2}) and (\ref{eq:6.10}), the eqs. (\ref{eq:6.11}) can be written in terms of scalar products and norms (all independent on $m$):
\begin{equation}\label{eq:6.16}
\lim_{r\rightarrow\infty}r\, e^{k_{n}^{\left(l\right)}r}\left(a_{r'}^{\left(l,m\right)}\psi_{0},\xi_{n,r}^{\left(l,m\right)}\right)=d_{n}^{\left(l\right)}\left(a_{r'}^{\left(l,m\right)}\psi_{0},\varphi_{n}^{\left(l,m\right)}\right)
\end{equation}   
and
\begin{equation}\label{eq:6.17}
\lim_{r\rightarrow\infty}r\, e^{k_{n}^{\left(l\right)}r}||\xi_{n,r}^{\left(l,m\right)}||=|d_{n}^{\left(l\right)}|.
\end{equation}   
The limit of eq. (\ref{eq:6.17}) implies the existence of numbers $b_{n}^{\left(l \right)}$ such that
\begin{equation}\label{eq:6.18}
r\, e^{k_{n}^{\left(l\right)}r}||\xi_{n,r}^{\left(l,m\right)}||\leq b_{n}^{\left(l \right)},\quad\forall r.
\end{equation}   
Of course eq. (\ref{eq:6.18}) is weaker than eq. (\ref{eq:6.17}). Now, we shall exhibit that the ultimate reason why the bound states 
$\varphi_{n}^{\left(l,m\right)}$ belong to $P^{\left(l,m\right)}\mathscr{H}$ is hidden into the eqs. (\ref{eq:6.16}) and 
(\ref{eq:6.18}) only.

In advance, let us estabilish a contact with \cite{VWH}. In our notations, and with a slight extension to $n'>0$, eq. (24) of \cite{VWH} reads
\begin{equation}\label{eq:6.19}
\sum_{\nu,\; n_{\nu}^{(l)}\neq0}\left(\varphi_{n'}^{\left(l,m\right)},a_{\nu}^{\left(l,m\right)}\psi_{0}\right)\frac{1}{n_{\nu}^{(l)}}\left(a_{\nu}^{\left(l,m\right)}\psi_{0},\varphi_{n}^{\left(l,m\right)}\right)=\delta_{n,n'}.
\end{equation}   
For $n'=n$, using the definition (\ref{eq:6.12}) of $P^{(l,m)}$, eq. (\ref{eq:6.19}) implies
\begin{equation}\label{eq:6.20}
\left(\varphi_{n}^{\left(l,m\right)},\left(1-P^{\left(l,m\right)}\right)\varphi_{n}^{\left(l,m\right)}\right)=0
\end{equation}   
which by the idempotency of the projection operator $1-P^{(l,m)}$ yields $\varphi_{n}^{(l,m)}=P^{(l,m)} \varphi_{n}^{(l,m)}$, i.e. the property that $\varphi_{n}^{(l,m)}$ belongs to the subspace $P^{(l,m)} \mathscr{H}$. Unfortunately the deduction of 
eq. (\ref{eq:6.19}) made in \cite{VWH} is not a true proof since it requires the exchange of a limit and a series which can be proved in no way. Nevertheless, the same result can be obtained rigorously using eqs. (\ref{eq:6.16}) and (\ref{eq:6.18}) alone. Really, eq. (\ref{eq:6.16}) implies by linearity
\begin{equation}\label{eq:6.21}
\lim_{r\rightarrow\infty}r\, e^{k_{n}^{\left(l\right)}r}\left(\varphi,\xi_{n,r}^{\left(l,m\right)}\right)=d_{n}^{\left(l\right)}\left(\varphi,\varphi_{n}^{\left(l,m\right)}\right)
\end{equation}   
for all the states $\varphi$ of the linear manifold spanned by the vectors $a_{r'}^{(l,m)} \psi_{0}$, $\forall r'$. This manifold is dense in $P^{\left(l,m\right)}\mathscr{H}$
by the mere definition of this subspace. In addition the vectors $r\, e^{k_{n}^{\left(l\right)}r}\xi_{n,r}^{\left(l,m\right)}$
form a bounded subset of $P^{\left(l,m\right)}\mathscr{H}$ due to eq. (\ref{eq:6.18}). Therefore, by Theor. 3 of \S 26 of \cite{akh}, eq.
(\ref{eq:6.21}) implies the weak limit
\begin{equation}\label{eq:6.22}
w-\lim_{r\rightarrow\infty}r\, e^{k_{n}^{\left(l\right)}r}\xi_{n,r}^{\left(l,m\right)}=d_{n}^{\left(l\right)}\varphi_{n}^{\left(l,m\right)}.
\end{equation}   
Thanks to the weak completeness of the Hilbert spaces, if $d_{n'}^{(l)} \neq 0$ $\forall n' \leq n$ we can deduce in the usual way from eq. (\ref{eq:6.22}) that $\varphi_{n}^{(l,m)}$ belongs to $P^{l,m} \mathscr{H}$. An analogous proof holds for the three-D case. The relation $\left(\varphi_{n'}^{\left(l,m\right)},P^{\left(l,m\right)}\varphi_{n}^{\left(l,m\right)}\right)=\left(\varphi_{n'}^{\left(l,m\right)},\varphi_{n}^{\left(l,m\right)}\right)=\delta_{n,n'}$ and the definition (\ref{eq:6.12}) of $P^{(l,m)}$ yield directly eq. (\ref{eq:6.19}). These results follow only from eqs. (\ref{eq:6.16}) and (\ref{eq:6.18}). Eq. (\ref{eq:6.17}) is the convergence of the norms of the vectors involved in eq. (\ref{eq:6.22}). If even this holds, eq. (\ref{eq:6.22}) implies the strong convergence
\begin{equation}\label{eq:6.23}
s-\lim_{r\rightarrow\infty}r\, e^{k_{n}^{\left(l\right)}r}\xi_{n,r}^{\left(l,m\right)}=d_{n}^{\left(l\right)}\varphi_{n}^{\left(l,m\right)}.
\end{equation}   
These considerations provide a precise mathematical support for the physical intuition that the V.W.H. method would be inexplicable if the bound states $\varphi_{n}^{(l,m)}$ would not live in $P^{(l,m)} \mathscr{H}$.

\section{Summary and concluding remarks}
In this paper we considered systems of identical particles interacting via a two-body potential possibly singular at finite distances and decreasing exponentially at infinity with arbitrary rate $R$ (see eqs. (\ref{eq:2.1}) and (\ref{eq:2.2})). Rotational invariance is not required. Our interest has been focused mainly on the nuclei, ignoring the Coulomb interaction from necessity. On the contrary the spin depending interactions have been disregarded only because the theorems of Sect. 4 require an upper bound for the ground state proved, in purely mathematical papers, for spinless particles only. If, as expected, this bound has a more general validity our proofs are consequently extended. Almost all the results of this paper follow from the exponential decay of the ground state $\psi_{0}$, of energy $E_{0}$, of an 
$(A+1)$-body nucleus when a nucleon is very far apart from the residual system. The center of mass motion is removed using the Jacobi coordinates $\mathbf{r}$ (of the nucleon) and $\boldsymbol{\xi}$ (\emph{internal} coordinates of the residual system). All the asymptotic behaviours are established in terms of limits in the direction $\widehat{ \mathbf{r}}$ of $\mathbf{r}$.

($a$) \emph{Asymptotic behaviour of the overlaps}. Let $\left(\varphi_{\lambda},a_{\mathbf{r}}\psi_{0}\right)$ be the overlap between $\psi_{0}$ and a bound or unbound eigenstate $\varphi_{ \lambda}$, of energy $\varepsilon_{ \lambda}$, of the Hamiltonian 
$H$ of the residual system. Let us settle $k_{\lambda}=2\mu_{0}\left(\varepsilon_{\lambda}-E_{0}\right)^{\frac{1}{2}}$, where 
$ \mu_{0}$ is the reduced mass of the distant nucleon. We obtained
\begin{equation}\label{eq:7.1}
\lim_{r\rightarrow\infty}r\, e^{k_{\lambda}r}\left(\varphi_{\lambda},a_{\mathbf{r}}\psi_{0}\right)=d_{\lambda}\left(\widehat{\mathbf{r}}\right),\quad\forall\varepsilon_{\lambda}<\varepsilon_{M},
\end{equation}   
where $d_{ \lambda}\left(\widehat{\mathbf{r}}\right)$, given by the second eq. (\ref{eq:5.2}), depends necessarily on $\widehat{ \mathbf{r}}$ and in general is not nought. In principle, cases of $d_{ \lambda}\left(\widehat{\mathbf{r}}\right)$ identically null are possible but in Nuclear Physics they should be considered rare anomalies. The energy $\varepsilon_{M}$, given by eq. (\ref{eq:5.1}), depends on $A$ in the light nuclei and decreases slowly for growing $R$ always remaining higher than the ground-state energy $\varepsilon_{0}$ of the residual nucleus. The behaviour of eq. (\ref{eq:7.1}), proposed first by T. Berggren \cite{berg} for an unspecified $d_{\lambda}$ in the case of finite-range potentials, up to now lacks a rigorous justification for interacting many-body systems. Our proof is rigorous but subject to an energy upper limit which is the best possible in our approach. We do not know whether a better result can be obtained in other ways but our conjecture is th!
 at some type of cut-off, progressively more severe for slowly decreasing interactions, has to exist.

($b$) \emph{Exponential decay of $\psi_{0}$}. It is deduced in the form
\begin{equation}\label{eq:7.2}
s-\lim_{r\rightarrow\infty}r\, e^{k_{n}r}\xi_{n,\mathbf{r}}=d_{n}\left(\widehat{\mathbf{r}}\right)\varphi_{n},\quad\forall\varepsilon_{n}<\min\left\{ \varepsilon_{C},\varepsilon_{M}\right\} ,
\end{equation}   
where $\varphi_{n}$ are the bound states of the residual nucleus, of energy $\varepsilon_{n}$, and
\begin{equation}\label{eq:7.3}
\xi_{n,\mathbf{r}}\equiv a_{\mathbf{r}}\psi_{0}-\sum_{n'=0}^{n-1}\left(\varphi_{n'},a_{\mathbf{r}}\psi_{0}\right)\varphi_{n'}.
\end{equation}   
This result is subject to the further restriction $\varepsilon_{n}$ lower than the threshold $\varepsilon_{C}$ of the continuous spectrum of the residual nucleus. This is purely due to the contribution of the unbound states of energy lower than $\varepsilon_{n}$. Two facts must be stressed. First, the limit of eq. (\ref{eq:7.2}) is understood in the sense of the strong convergence in the Hilbert space $\mathscr{H}$ of the states of the residual nucleus. Pointwise limits are useless here. Second, eqs. (\ref{eq:7.1}) and (\ref{eq:7.2}) are a strong way to express the exponential decay of $\psi_{0}$. As a matter of fact they imply
\begin{equation}\label{eq:7.4}
s-\lim_{r\rightarrow\infty}
\left[a_{\mathbf{r}}\psi_{0}
-\sum_{n'=0}^{n}d_{n'}\left(\widehat{\mathbf{r}}\right)\frac{e^{-k_{n'}r}}{r}\varphi_{n'} \right]=0,
\end{equation}
i.e. the most common way of expressing the asymptotic behaviours in Physics, but this   
 is a weaker property insufficient for the purposes of this paper. Eq. (\ref{eq:7.2}) is rich in consequences, all subject to the condition 
$\varepsilon_{n} < \min\left\{ \varepsilon_{C},\varepsilon_{M}\right\} $.

($b.1$) It implies the weak convergence
\begin{equation}\label{eq:7.5}
\lim_{r\rightarrow\infty}r\, e^{k_{n}r}\left(\varphi,\xi_{n,\mathbf{r}}\right)=d_{n}\left(\widehat{\mathbf{r}}\right)\left(\varphi,\varphi_{n}\right),\quad\forall\varphi\in\mathscr{H},
\end{equation}   
which for $n=0$ and a suitable choice of $\varphi$ establishes the asymptotic behaviour of many single-hole quantities, as the hole Green's function at complex energies and the related hole Hamiltonian. For $n\geq0$, choosing $\varphi=a_{\mathbf{r}'}\psi_{0}$, eq. 
(\ref{eq:7.5}) yields the asymptotic behaviour of the (subtracted) density matrices
\begin{alignat}{1} \label{eq:7.6}
K_{n}\left(\mathbf{r},\mathbf{r}'\right) &
=\left(a_{\mathbf{r}'}\psi_{0},\xi_{n,\mathbf{r}}\right)
=K \left(\mathbf{r},\mathbf{r}'\right)
-\sum_{n'=0}^{n-1}\left(a_{\mathbf{r}'}\psi_{0},\varphi_{n}\right)\left(\varphi_{n},a_{\mathbf{r}}\psi_{0}\right),\nonumber \\
K_{0}\left(\mathbf{r},\mathbf{r}'\right) &
= K \left(\mathbf{r},\mathbf{r}'\right) \equiv (a_{\mathbf{r}'}\psi_{0},a_{\mathbf{r}}\psi_{0})
\end{alignat}   
which is
\begin{equation}\label{eq:7.7}
\lim_{r\rightarrow\infty}r\, e^{k_{n}r}K_{n}\left(\mathbf{r},\mathbf{r}'\right)=d_{n}\left(\widehat{\mathbf{r}}\right)\left(a_{\mathbf{r}'}\psi_{0},\varphi_{n}\right).
\end{equation}   

($b.2$) Eq. (\ref{eq:7.2}) implies the corresponding limit of the related square norms $|| \xi_{n, \mathbf{r}}||^{2}$ which coincide with the (subtracted) nuclear densities $ \rho_{n} (\mathbf{r})=K_{n}(\mathbf{r}, \mathbf{r})$:
\begin{equation}\label{eq:7.8}
\lim_{r\rightarrow\infty}r^{2}e^{2k_{n}r}\rho_{n}\left(\mathbf{r}\right)=|d_{n}\left(\mathbf{r}\right)|^{2}.
\end{equation}
Eqs. (\ref{eq:7.7}) and (\ref{eq:7.8}) are the conceptual basis of the V.W.H. method \cite{VWH} developed here in Sect. 6.

($b.3$) \emph{Hilbert properties of the bound states of the nuclei}. Let us suppose that the coefficients 
$d_{n}\left(\widehat{\mathbf{r}}\right)$ are nonnull, even in different directions $\widehat{\mathbf{r}}$. Therefore in successive steps eq. (\ref{eq:7.2}) implies that the bound eigenstates $\varphi_{n}$ belong to the subspace $P \mathscr{H}$, range of the projection operator 
\begin{equation}\label{eq:7.9}
P\varphi= \sum_{{\nu\atop n_{\nu} \neq 0}} \left(a_{\nu}\psi_{0},\varphi\right)\frac{1}{n_{\nu}}a_{\nu}\psi_{0},
\quad a_{\nu}\psi_{0} \equiv \int\overline{u}_{\nu}\left(\mathbf{r}\right)a_{\mathbf{r}}\psi_{0} d \mathbf{r},
\end{equation}   
where $u_{ \nu}$ and $n_{ \nu}$ are the natural orbitals and the related occupancies of the unsubtracted density matrix $K$. We have deduced that $\varphi_{n}$ belongs to $P \mathscr{H}$ even under conditions weaker than eq. (\ref{eq:7.2}): eq. (\ref{eq:7.7}), with $d_{n}( \widehat{ \mathbf{r}}) \neq 0$, and the boundness of the function $r^{2} e^{2 k_{n} r} \rho_{n}(\mathbf{r})$, consequence of eq. (\ref{eq:7.8}). This proof, given at the end of Sect. 6 in the partial-wave context, shows that the property $\varphi_{n} \in P \mathscr{H}$ is congenital in the V.W.H. method. This property, which was the initial motivation of this paper, means that every nucleus daughter of a stable nucleus with an extra nucleon has bound eigenstates $\varphi_{n}$ belonging to the subspace $P \mathscr{H}$, range of the projection operator $P$ associated with the ground state of the parent nucleus. This property characterizes strongly the eigenstates $\varphi_{n}$ since $P \mathscr{H}$ is, so to!
  say, a rather small subspace of $\mathscr{H}$: one could find infinitely many subspaces of $\mathscr{H}$ of infinite dimension which are orthogonal to $P \mathscr{H}$. In the single-hole theory, developed in \cite{boffi} and \cite{mah}, this restriction has a meaningful consequence which could not be foreseen at that time. The dynamical part of the single-hole Hamiltonian $H^{(h)}$ defined in eq. (\ref{eq:2.10}) is not null (differently from the uncorrelated case), but it has no role in determining the single-hole overlaps. This displays a deep difference from the Hamiltonian for the self-energy whose dynamical part is essential to determine the single-hole overlaps with the bound eigenstates $\varphi_{n}$ \cite{capma}. Note that the effective energy independence of $H^{(h)}$ is a peculiarity of the hole case only. Nothing of the sort can be found in the Feshbach's Hamiltonian (\cite{Feshbach}, \cite{mah}) which provides the single-particle overlaps.

One could raise the following objection to the result $\varphi_{n} \in P \mathscr{H}$. This subspace should be composed of states of the form $a_{f} \psi_{0}$, where $f$ is a square summable function of $\mathbf{r}$ (single-hole states). Therefore the eigenstates
$\varphi_{n}$ could not belong to $P \mathscr{H}$ because they can be single-hole states only in the uncorrelated (or partially correlated) systems. This objection is unfounded since it is based on the idea of a space $P \mathscr{H}$ structured as in the independent particle model. Really the study of this subspace for the correlated systems performed in Subsect. 5.6 stresses many substantial differences. In the independent particle model the dimension of $P \mathscr{H}$ is $A+1$ and all its elements are single-hole states. On the contrary in the fully correlated systems its dimension is infinite. This has two consequences. First, the occupation numbers 
$n_{ \nu}$ accumulate at zero, which makes problematic the convergence of some series involving $\frac{1}{n_{\nu}}$. Second, the range of the unsubtracted density matrix is never a closed manifold. It is just this property that allows $P \mathscr{H}$ to contain generalized single-hole states, i.e. strong limits of single-hole states which cannot be represented in the form $a_{f} \psi_{0}$ for any square summable $f$. One can construct a lot of these states. Surprisingly even the states $a_{ \mathbf{r}} \psi_{0}$ are of this type. It is only in the finite dimensional case that they are single-hole states. An argument developed at the end of Subsect. 5.6 provides the indirect evidence that the bound eigenstates $\varphi_{n}$ are generalized single-hole states.

For sake of generality and to avoid too many indices we have considered potentials without spherical symmetry. In Sect. 6 we treated central potentials to extend our approach to degenerate eigenvalues. By proving eq. (\ref{eq:7.2}), specialized for the partial-wave components $\xi^{(l,m)}_{n,r}$ of $\xi_{n, \mathbf{r}}$, we have obtained the rigorous justification of the V.W.H. method. Moreover the Hilbert properties of the bound eigenstates $\varphi_{n}^{(l,m)}$ of the residual nucleus have been distinguished introducing, for every pair of quantum numbers $l$ and $m$, more specialized projection operators $P^{(l,m)}$ with ranges $P^{(l,m)} \mathscr{H}$ containing the vectors $\varphi_{n}^{(l,m)}$. These ranges are mutually orthogonal subspaces of $P \mathscr{H}$ which is their direct sum.

In this paper the proof of the property $\varphi_{n} \in P \mathscr{H}$ is subject to three conditions: ($i$) $\varphi_{n}$ is not embedded in the continuum, ($ii$) its energy $\varepsilon_{n}$ is lower than a maximum energy $\varepsilon_{M}$ and 
($iii$) the coefficient $d_{n}( \widehat{ \mathbf{r}})$ is not identically null. The exclusion ($i$) is essential because the bound states embedded in the continuum can have nonnull components both in $P \mathscr{H}$ and in its orthogonal complement 
$Q \mathscr{H}$, as shown at the end of Subsect. 5.4. The restriction ($ii$) only concerns the exponential decay of $\psi_{0}$. No reason prevents $\varphi_{n}$ from belonging to $P \mathscr{H}$ by other reasons. The case of $d_{n}( \widehat{ \mathbf{r}})$ identically null is a rare anomaly, at least in Nuclear Physics, which probably does not allow the property 
$\varphi_{n} \in P \mathscr{H}$. Work is in progress to develope a new approach which investigates the ultimate physical reason of the property $\varphi_{n} \in P \mathscr{H}$ and the meaning of the anomaly 
$d_{n}( \widehat{ \mathbf{r}})=0$, $\forall \widehat{ \mathbf{r}}$.

\section*{Acknowledgements}
I thank S. Boffi, C. Giusti, F.D. Pacati and A. Rimini for helpful discussions.

\end{document}